\documentclass[prx,aps,amsmath,amssymb,showpacs,twocolumn]{revtex4-1}
\usepackage{amssymb,graphicx,float,dcolumn,bm,subfigure,color}
\usepackage[titletoc,title]{appendix}
\usepackage[dotinlabels]{titletoc}
\usepackage{cancel}
\usepackage{subfigure}
\usepackage{mathrsfs}
\usepackage[mathscr]{euscript}
\usepackage{hyperref}
\usepackage[normalem]{ulem}
\newcommand{\be}{\begin{equation}}
\newcommand{\ee}{\end{equation}}

\newcommand{\ba}{\begin{eqnarray}}
\newcommand{\ea}{\end{eqnarray}}

\renewcommand{\vec}[1]{\mbox{\boldmath$#1$}}

\newcommand\eg{\emph{e.g.}~}
\newcommand\ie{\emph{i.e.}~}
\newcommand{\lm}{\ell_B}

\newcommand{\sh}{\mathcal{S}}

\def\beq{\begin{eqnarray}}

\newcommand{\redTwo}{{\color{red}{2}}}
\newcommand{\PLLL}{{P_{\rm LLL}}}
\def\eeq{\end{eqnarray}}
\renewcommand{\i}{i}

\newcommand{\elliptic}[5][\scriptstyle]{\vartheta\left[\begin{array}{c}{{#1 #2}}\\{#1 #3}\end{array}\right]\left(#4\middle|#5\right)}

\makeatletter
\newcommand*{\rom}[1]{\expandafter\@slowromancap\romannumeral #1@}
\makeatother

\begin{document}

\title{Hall Viscosity of Composite Fermions}

\author{Songyang Pu$^1$, Mikael Fremling$^2$, J. K. Jain$^1$}
\affiliation{$^1$Department of Physics, 104 Davey Lab, Pennsylvania State University, University Park, Pennsylvania 16802, USA.\\
$^2$Institute for Theoretical Physics, Center for Extreme Matter and Emergent Phenomena,
Utrecht University, Princetonplein 5, 3584 CC Utrecht, the Netherlands
  }

\date{\today}

\begin{abstract} 
Hall viscosity, also known as the Lorentz shear modulus, has been proposed as a topological property of a quantum Hall fluid.
Using a recent formulation of the composite fermion theory on the torus,
we evaluate the Hall viscosities for a large number of fractional quantum Hall states at filling factors of the form $\nu=n/(2pn\pm 1)$,
where $n$ and $p$ are integers, from the explicit wave functions for these states.
The calculated Hall viscosities $\eta^A$ agree with the expression $\eta^A=(\hbar/4) {\cal S}\rho$,  
where $\rho$ is the density and ${\cal S}=2p\pm n$ is the ``shift'' in the spherical geometry. 
We discuss the role of modular covariance of the wave functions,
projection of the center-of-mass momentum, and also of the lowest-Landau-level projection.
Finally, we show that the Hall viscosity for $\nu={n\over 2pn+1}$ may be derived analytically from the microscopic wave functions,
provided that the overall normalization factor satisfies a certain behavior in the thermodynamic limit.
This derivation should be applicable to a class of states in the parton construction,
which are products of integer quantum Hall states with magnetic fields pointing in the same direction.
\end{abstract}

\maketitle
\section{Introduction}

The extreme precision of the quantization of the Hall resistance in the integer quantum Hall effect~\cite{Klitzing80} led to a topological interpretation in terms of Chern numbers~\cite{Thouless82}.  The fractional quantum Hall effect \cite{Tsui82},
which emerges as a result of strong interactions, also motivated quantities that have a topological origin.
These include the fractional charge for the excitations~\cite{Laughlin83} and the vorticity of composite fermions that is reflected through an effective magnetic field \cite{Jain89,Jain90,Jain07}.
According to a general topological classification based on the Chern-Simons theory \cite{Wen92,Wen92a,Wen95},
the fractional quantum Hall ground states are characterized not only by the electromagnetic response,
\eg the Hall conductance,
but also by the geometrical response, which can also be topological.
An example of this is the Hall viscosity, the topic of this article.

To understand the Hall viscosity as a geometrical response, let us consider applying a small deformation to a fluid.
The small deformation is represented by the strain tensor $u_{ij}$ and the strain-rate tensor $\dot{u}_{ij}={du_{ij}\over dt}$,
with $u_{ij}=(\partial_i u_j+\partial_j u_i)/2$, where $u_i(\vec{r})$ is the displacement at $\vec{r}$ in the $i$th direction.
The stress tensor $\sigma_{ij}$ induced by this small deformation is given by:
\be
\sigma_{ij}=
\sum_{k,l}\lambda_{ijkl}u_{kl}
+\sum_{k,l}\eta_{ijkl}\dot{u}_{kl}.
\ee
Of the two rank-four tensors that describe the response to the deformation,
the first one $\lambda_{ijkl}$ is called the elastic modulus tensor,
which measures the fluid's resistance to elastic deformation.
The viscosity tensor $\eta_{ijkl}$ describes the fluid's resistance to being deformed at a given rate.
Because both the stress tensor and strain-rate tensor are symmetric under the exchange $i\leftrightarrow j$ or $k\leftrightarrow l$,
the viscosity tensor also has the symmetry property $\eta_{ijkl}=\eta_{jikl}=\eta_{ijlk}$. Considering the exchange $ij\leftrightarrow kl$,
the viscosity tensor can be further decomposed into a symmetric component $\eta^S_{ijkl}=(\eta_{ijkl}+\eta_{klij})/2$ and an antisymmetric component $\eta^A_{ijkl}=(\eta_{ijkl}-\eta_{klij})/2$.
The symmetric component is associated with the energy dissipation of the deformation,
while the antisymmetric component is nondissipative. As a result of the Onsager relation in thermal physics,
the antisymmetric component survives only when the deformation is irreversible,
\ie the time-reversal symmetry is broken, such as in a quantum Hall system.
In a two-dimensional isotropic system, the antisymmetric component is given by~\cite{Avron95}
\be
\eta^A=\eta^A_{1112}=\eta^A_{1222}.
\ee
It is called the Hall viscosity in a quantum Hall system, and is also referred to as the Lorentz shear modulus~\cite{Tokatly07b,Tokatly09}.

\begin{figure}[t]
  \includegraphics[width=\columnwidth]{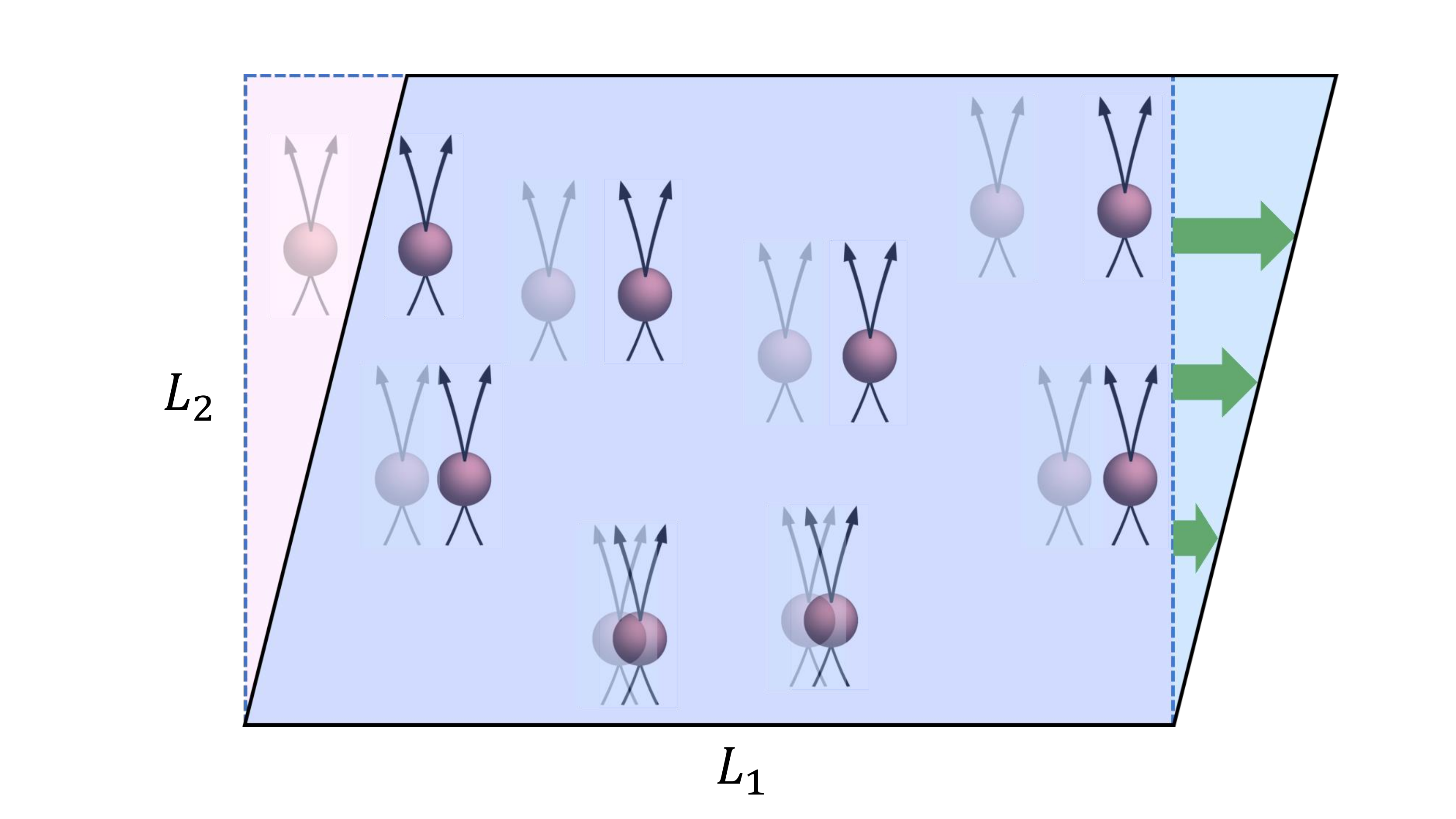} 
  \caption{
    An schematic illustration of deforming a torus. The coordinate of a composite fermions is given by $z_i=x_i+iy_i=L_1\theta_{1,i}+L_2\theta_{2,i}$. With the deformation of torus, the physical coordinate $(x_i,y_i)$ changes while the reduced coordinate $(\theta_{1,i},\theta_{2,i})$ is fixed.
  }
  \label{Fig0}
\end{figure}

The Hall viscosity is a local bulk property of a fluid and should therefore be present (and measurable) independently of the geometry.
However, for its theoretical evaluation it has proven particularly fruitful to consider a system with periodic boundary conditions, \ie a torus, as shown in Fig.~\ref{Fig0}.
One may see this by noting that by adiabatically changing the torus geometry, one effectively simulates a uniform infinitesimal strain rate in the fluid.

Avron, Seiler, and Zograf~\cite{Avron95} (ASZ) considered the fluid on a torus defined by two sides $L_1$ and $L_2=L_1\tau$ in the complex plane.
Here  $\tau=\tau_1+i\tau_2$ is the modular parameter that specifies the geometry of the torus.
In what follows, we choose $L_1$ to be real and $\tau_2>0$, in which case the area of the torus is 
$V=L_1^2\tau_2$.
The number of flux quanta passing through the torus, $N_\phi$, is quantized to be an integer.
(Here the flux quantum is defined as $\phi_0=h/e$.)
The filling factor is defined as $\nu=N/N_\phi$, where $N$ is the number of electrons.
It was shown by ASZ that the antisymmetric Hall viscosity $\eta^A$ is related to the Berry curvature $\mathcal{F}_{\tau_1,\tau_2}$ in $\tau$ space (\ie the space of torus geometries) as 
\be
\label{berry-curv}
\eta^A=-{\hbar \tau_2^2  \over V}\mathcal{F}_{\tau_1,\tau_2},
\ee
where 
\be
\label{BC}
\mathcal{F}_{\tau_1,\tau_2}=-2{\rm Im}\bigg\langle {\partial \Psi \over \partial \tau_1}\bigg|{\partial \Psi \over \partial \tau_2}\bigg\rangle.
\ee
The wave function $\Psi$ is a function of $\tau_1$, $\tau_2$ and particle coordinates $z_i=x_i+iy_i=L_1\theta_{1,i}+L_2\theta_{2,i}$,
where $\theta_{1,i}, \theta_{2,i}\in [0,1)$ are called the reduced coordinates.
The particles' physical coordinates $x_i$ and $y_i$  change with the deformation of the geometry but the reduced coordinates do not; the partial derivatives with respect to $\tau_1$ and $\tau_2$ in Eq.~\ref{BC} are evaluated while keeping the reduced coordinates $\theta_{1,i}$ and $\theta_{2,i}$ invariant.
Ref.~\onlinecite{Avron95} further showed that in the lowest Landau level (LLL),
the Berry curvature for a single particle in any orbital is ${1\over 4\tau_2^2}$,
implying that the Hall viscosity for a filled LLL is ${\hbar \over 4}{N \over V}$.
Note that ${1\over \tau_2^2}d\tau_1\wedge d\tau_2$ is the volume form in $\tau$ space,
implying that the Berry curvature is a constant (${1\over 4}$) times the factor of the volume form.

ASZ further showed~\cite{Avron95} that the integral of $\mathcal{F}_{\tau_1,\tau_2}$ in Eq.~\ref{BC} in the fundamental domain of the $\tau$ plane gives the Hall viscosity,
just as the integration of Berry curvature in the first Brillouin zone of $k$ space returns the Hall conductivity.
(Roughly speaking, the fundamental domain is the subset of all points in the $\tau$ plane which are not related by modular transformations \cite{Gunning62}.
More details on the modular transformations are given in Sec.~\ref{sec:wfns} and Appendix~\ref{app:mod_vis}.)

L\'evay \cite{Levay95} showed that the Berry curvature for a single particle in the $n$th Landau level (with $n=0$ for LLL) is given by
\be
\mathcal{F}_{\tau_1,\tau_2}=-{1\over 2}\left(n+{1\over 2}\right){1\over \tau_2^2}.
\ee
The Hall viscosity for an integer number of filled Landau levels (LLs) can then be straightforwardly derived by adding the contributions of all single particles (which is valid for a Slater determinant state) to give
 \be
\label{intfill}
 \eta^A_n=n{\hbar \over 4}{N \over V}.
 \ee
 
The calculation of the Hall viscosity becomes more complicated for fractional quantum Hall (FQH) states, which do not have a single Slater determinant form.
For the Laughlin state at filling ${1\over m}$\cite{Laughlin83},
the Hall viscosity was shown to be $\eta^A=m{\hbar \over 4}{N \over V}$ by Tokatly and Vignale~\cite{Tokatly09}, and by Read~\cite{Read09} by exploiting the plasma analogy.
Read~\cite{Read09} also showed that for FQH states that can be expressed as conformal blocks in a conformal field theory,
for which a generalized plasma analogy exists, the Hall viscosity satisfies
\be
 \label{hall visc}
 \eta^A=\sh{\hbar \over 4}{N \over V}.
 \ee
Here $\sh$ is the so-called ``shift", \ie the offset of flux quanta needed to form a ground state on a sphere:
 \be
 \sh={N\over \nu}-N_\phi
 \ee

The shift in the spherical geometry is a manifestation of the orbital spin,
introduced by Wen and Zee~\cite{Wen92} to describe the coupling between the orbital motion and the curvature of space.
This topological quantum number is quantized and robust within a topological phase,
and can thus distinguish between different topological phases.
The Hall viscosity is by extension believed to be a topological quantum number for a given FQH state.
In some sense, it is a manifestation of the orbital spin through a transport coefficient.
Read and Rezayi \cite{Read10} discussed the connection between Hall viscosity and the orbital spin \ie Eq.~\ref{hall visc},
along with the robustness of Hall viscosity within a topological phase,
by noting that the commutator of distinct shear operations is a rotation.
 
The Hall viscosity of the Laughlin and the paired Hall states has also been extracted by Lapa {\em et al.}~\cite{Lapa18,Lapa18b} from the matrix models for these states.
Cho, You, and Fradkin~\cite{Cho14} showed that Eq.~\ref{hall visc} can be derived for the Jain states from the effective Chern-Simons field theory. 
In this paper, our aim is to obtain the Hall viscosity of the  Jain states at filling fractions
\be 
\nu={n\over 2pn\pm 1},
\ee
directly from the explicit microscopic wave functions. 
According to Eq.~\ref{hall visc}, we expect 
\be
 \eta^A={\hbar N\sh \over 4V}={\hbar N (2p \pm n) \over 4V}\;.
\label{hall visc2}
\ee
An important step in this direction was taken by Fremling, Hansson, and Surosa~\cite{Fremling14},
who obtained the Hall viscosity for the 2/5 state using both the exact Coulomb wave functions,
and a trial wave function obtained from conformal field theory (CFT) correlation functions.
While approaching Eq.~\ref{hall visc} with increasing system size,
the results showed significant variation with the aspect ratio $\tau_2$, and also with the number of particles, presumably because the calculations were limited to small systems (10 particles or fewer).

\emph{Main objectives and results}:
We briefly describe the two main objectives for the present work, along with a summary of the results.
First, a recent work~\cite{Pu17} has demonstrated how to construct the Jain wave functions on the torus for large systems,
which should allow a treatment of many fractions of the form $\nu=n/(2pn+1)$ and also an estimation of the thermodynamic limit.
Our explicit numerical evaluation of the Hall viscosity for many FQH states produces values consistent with Eq.~\ref{hall visc2} in the thermodynamic limit.
The second objective for our work is to seek an analytical evaluation of the Hall viscosity for general FQH states, starting from the microscopic wave function.
The unprojected wave function for the state at $\nu=n/(2pn+1)$ is given by $Z \Psi_n \Psi_1^{2p}$,
where $\Psi_n$ is the normalized wave function of $n$ filled Landau levels and $Z$ is an overall normalization factor.
We prove that the Hall viscosity of this wave function is equal to the sum of the Hall viscosities of the individual factors of $\Psi_n$ and $\Psi_1$ (given in Eq. ~\ref{intfill}), producing Eq.~\ref{hall visc2},
provided that we make an assumption regarding the behavior of $Z$ in the thermodynamic limit. 
We further show that the Hall viscosity of the LLL projected wave function $Z'\PLLL \Psi_n \Psi_1^{2p}$ (where $\PLLL$ is the LLL projection operator) is also given by the sum of the Hall viscosities of the individual factors of $\Psi_n$ and $\Psi_1$,
provided that we make an assumption regarding the behavior of $Z'$ in the thermodynamic limit.
Numerical calculations provide support to the assumption regarding the normalization factors. 

The connection between the Hall viscosity and the Berry curvature, as encapsulated in Eq.~\ref{berry-curv},
relies on the adiabatic theorem, \ie on the presence of a gap.
In particular, it applies to the incompressible Jain states.
For the composite-fermion (CF) Fermi-sea states at even-denominator fillings,
Eq.~\ref{berry-curv} is no longer valid due to the absence of a gap, and therefore, our approach does not apply to these states.
The Hall viscosity is also not expected to be quantized for the CF Fermi sea.

We discuss the modular covariance of the Jain states, which is an important requirement from legitimate wave functions in the torus geometry as well as crucial for the evaluation of the Hall viscosity.
We also find that projection of the wave function onto a well-defined center-of-mass momentum sector produces a better quantized Hall viscosity for finite $N$,
although in the thermodynamic limit the quantized value is obtained independently of the center-of-mass momentum projection.
In Appendices \ref{App:MomProj} and \ref{App:Momentum_Mixing} we discuss this issue and show that momentum projection is a sub-leading effect in the thermodynamic limit.

We briefly mention proposals for measuring the Hall viscosity experimentally. 
Haldane~\cite{Haldane09} suggested that it is directly related to the stress caused by an inhomogeneous electric field.
Following this direction, Hoyos and Son~\cite{Hoyos12} and Bradlyn,
Goldstein, and Read~\cite{Bradlyn12} have shown the Hall viscosity can be extracted from the wave-vector-dependent contribution to the Hall conductance. 
Based on this idea, Ref.~\cite{Delacretaz17} proposed that the Hall viscosity in the quantum Hall regime can be extracted in a pipe-flow setup,
by measurement of the  electroscalar potential close to the point contacts where current is injected.
On the other hand,
Refs.~\cite{Alekseev16,Scaffidi17,Pellegrino17,Steinberg58,Ganeshan17} discussed the Hall viscosity and how to measure it in the hydrodynamic regime using semi-classical theory.
Such theoretical proposals have been applied to graphene \cite{Berdyugin18} and GaAs quantum wells \cite{Gusev18}.
There, a negative magnetoresistance and suppression of Hall resistivity were observed as a manifestation of Hall viscosity in the hydrodynamic regime.
According to the analysis of Refs.~\cite{Alekseev16,Scaffidi17},
the Hall viscosity in the hydrodynamic regime approaches the values found in the integer quantum Hall regime,
in the limit $B\rightarrow \infty$, and makes the same contribution to Hall conductivity as found in Ref.~\cite{Hoyos12}.

The plan for the remainder of the article is as follows:
In Sec.~\ref{sec:wfns} we introduce the wave functions we use to evaluate Hall viscosities.
These were originally constructed in Ref.~\onlinecite{Pu17},
but here we use a slightly different (though equivalent) form,
using the so-called $\tau$ gauge, which is crucial for our analytical proof.
(Both the symmetric and the $\tau$ gauges are equally good for Monte Carlo evaluations.)
In Sec.~\ref{sec:Analytical_proof}, we prove that the Hall viscosity at $\nu={n\over 2pn+1}$ is given by Eq.~\ref{hall visc2},
provided that we assume that the overall normalization factor of the product wave functions has a certain behavior in the thermodynamic limit; this assumption is tested numerically.
In Sec.~\ref{numerical}, we numerically evaluate the Hall viscosities at fillings
$\nu=2/5$, $3/7$, $2/9$, $2/3$, $3/5$, and $2/7$ and find that the thermodynamic limit of the result is consistent with Eq.~\ref{hall visc2}.
We also find that both the projected and the unprojected wave functions produce the same Hall viscosity,
supporting the notion that it is a topological quantity.
To conclude, the key results of this work are the derivation of Read's formula Eq.~\ref{hall visc} through CF wave functions, presented in Sec.~\ref{sec:Analytical_proof},
and the numerical evaluations of Eq.~\ref{berry-curv} for various Jain states, presented in Sec.~\ref{numerical}.

\section{Composite Fermions on the Torus}
\label{sec:wfns}

\subsection{Modular covariance of wave functions}

As mentioned in the previous section, the torus is topologically equivalent to a periodic lattice spanned by a parallelogram.
There are an infinite number of choices of basis vectors that span the same lattice.
New basis vectors can be obtained from old basis vectors by the transformation $\bigl(
\begin{smallmatrix}
L_2'\\
L_1'
\end{smallmatrix}
\bigr)
=
\bigl(
\begin{smallmatrix}
a &b\\
c &d
\end{smallmatrix}
\bigr)
\bigl(
\begin{smallmatrix}
L_2\\
L_1
\end{smallmatrix}
\bigr)
$,
where $a,b,c,d\in\mathbb{Z}$, $ad-bc=1$. In terms of the modular parameter $\tau=L_2/L_1$,
this corresponds to the transformation $\tau\to\tau^\prime=(a\tau+b)/(c\tau+d)$.
Since changing the signs of all elements does not produce a new transformation,
the matrices $\bigl(\begin{smallmatrix}a&b\\c&d
\end{smallmatrix}\bigr)$
form the $SL(2,\mathbb{Z})/\mathbb{Z}_2$ group.
This group is spanned by two modular transformations $T: \tau\rightarrow \tau+1$ and
$S: \tau\rightarrow -{1/\tau}$, which, in the matrix representation, correspond to
$T=  \bigl(\begin{smallmatrix}1&1\\0&1
\end{smallmatrix}\bigr)$
and $S=  \bigl(\begin{smallmatrix}0&1\\-1&0
\end{smallmatrix}\bigr)$.
We will define the ``physical" coordinates $(x,y)$ (also expressed as the complex number $z=x+iy$) with reference to the Cartesian axis. These coordinates span the whole complex plane and thus do not depend on the choice of the lattice vectors,
\ie remain unaltered upon a modular transformation. 

The wave functions in general depend on the modular parameter $\tau$
and thus are not necessarily invariant under modular transformations $\tau\rightarrow \tau+1$ and $\tau \rightarrow -1/\tau$. However,
because a modular transformation commutes with the Hamiltonian (which is invariant under a modular transformation), all sets of degenerate states,
and by extension their observables, must be closed under modular transformations. That is,
a modular transformation may only mix states that are degenerate in energy.
Degenerate sets of wave functions that satisfy this property are said to be modular covariant.

There are $2pn\pm 1$ degenerate ground state wave functions at $\nu=n/(2pn\pm 1)$. 
The set of exact ground state wave functions is of course closed under modular transformation.
However, a given trial wave function does not necessarily satisfy the property of modular covariance.
Explicit calculations in Ref.~\cite{Fremling14} have demonstrated that the Hall viscosity is sensitive to modular covariance of the wave function,
and a well-defined value is obtained only for wave functions that are modular covariant.
It is therefore important to ensure that the wave function we are using to calculate the Hall viscosity is modular covariant.
The modular covariance of the Jain composite-fermion (CF) wave functions can be demonstrated quite straightforwardly in the following fashion ~\cite{Fremling19}. 

First of all, since the wave function $\Psi_n$ of $n$ filled LLs is non-degenerate,
it is modular {\em invariant}; \ie it remains unaltered under modular transformations.
(For one filled LL, this can be proved by construction of single-particle coherent wave functions,
using Haldane's modified sigma functions, that are modular invariant~\cite{Fremling19}.)
It therefore follows that the unprojected product wave function $\Psi^{\rm unproj}_{n\over 2pn\pm 1}\sim\Psi_n\Psi_1^{2p}$ is also modular invariant. 

It further follows that the LLL-projected wave function $\Psi_{n\over 2pn\pm 1}=P_{\rm LLL}\Psi^{\rm unproj}_{n\over 2pn\pm 1}$
is modular invariant as well. This is easiest to see when $P_{\rm LLL}$ is defined as what is known as the direct projection.
This $P_{\rm LLL}$ is explicitly modular invariant, as evident from its definition
${\PLLL}=\prod_{i=1}^N\prod_{n=1}^\infty(1-a_i^\dagger a_i/n)$.
Here $a_i$ and $a_i^\dagger$ are the LL lowering and raising operators,
with $a_i^\dagger a_i$ measuring the LL index of the $i$th particle.
The situation is more subtle for a different projection, introduced by Pu,
Wu, and Jain (PWJ)~\cite{Pu17} as the torus generalization of the Jain-Kamilla projection~\cite{Jain97,Jain97b}.
The PWJ projection will be used below in our calculation below,
as it allows treatment of large systems in the torus geometry. The PWJ projection cannot be represented by an operator,
but it is possible to show that the PWJ-projected wave functions are also modular covariant; we refer to the work by Fremling~\cite{Fremling19} for a detailed proof.
Interestingly the condition for modular covariance of the wave function is the  same as that for the validity of the PWJ projection; namely, that the states are proper states,
where proper states correspond to configurations for which there are no unoccupied CF-orbitals directly beneath an occupied CF-orbital~\cite{Pu17}.
(The PWJ projection does not preserve the quasi-periodic boundary conditions for non-proper states.)

Let us next address the fact that the ground state is not unique but there are $2pn\pm 1$ degenerate ground states at $\nu=n/(2pn\pm 1)$,
which can be chosen as eigenstates of the center-of-mass momentum $M$. As shown in the Appendix \ref{App:MomProj},
these can be obtained as $\Psi_{n\over 2pn\pm 1}^{(M)}=\mathcal{P}_M\Psi_{n\over 2pn\pm 1}$, where $\mathcal{P}_M$ is the momentum projection operator. Under a modular transformation, we get \cite{Fremling19}
\ba
&&\Psi_{n\over 2pn\pm 1}^{(M)}=\mathcal{P}_M\Psi_{n\over 2pn\pm 1} \nonumber\\
&&\quad\quad\rightarrow \mathcal{P}'_M\Psi_{n\over 2pn\pm 1}=\sum_{M'}K_{M,M'}\Psi_{n\over 2pn\pm 1}^{(M')},
\label{KMM}
\ea
where $K_{M,M'}$ is a unitary matrix. In other words, the set of $2pn\pm 1$ degenerate ground states $\{\Psi_{n\over 2pn\pm 1}^{(M)}\}$ is closed under modular transformation.
This proves the modular covariance of the Jain CF wave functions. 

Finally, let us consider a translationally invariant physical operator $\hat{O}$, which must be independent of the center-of-mass momentum,
\ie must satisfy $\langle\Psi^{(M)}| \hat{O}|\Psi^{(M')}\rangle=O \delta_{M,M'}$.
It follows that the expectation value $O$ remains invariant under modular transformation provided that $\Psi^{(M)}$ transforms according to Eq.~\ref{KMM}.

The Hall viscosity, given by $\hat\eta\propto\overleftarrow{\partial}_{\tau_{1}}\tau_{2}^{2}\overrightarrow{\partial}_{\tau_{2}}$,
is modular invariant. That is demonstrated in Appendix~\ref{app:mod_vis}.

We note that the modular covariance of the wave functions obtained in the parton construction~\cite{Jain89b} also follows as above,
because these wave functions are also products of integer quantum Hall states. 

\subsection{Wave functions in the ``$\tau$ gauge"}
\label{tau wf}
A proper gauge choice can be important.
The wave functions for general FQH states at $\nu=n/(2pn\pm 1)$ in Ref.~\cite{Pu17} and for the CF Fermi sea in Refs.~\onlinecite{Shao15,Wang19,Geraedts18,Pu18} were first constructed in the symmetric gauge.
This was crucial for implementing LLL projection using the PWJ method.
Fremling~\cite{Fremling19} demonstrated that these wave functions are modular covariant, \ie satisfy Eq.~\ref{KMM}.
Here we use the Jacobi theta functions with rational characteristics, given by\cite{Mumford07}
\be
\elliptic[\displaystyle]abz\tau=\sum_{n=-\infty}^{\infty}e^{i\pi \left(n+a\right)^2\tau}e^{i2\pi \left(n+a\right)\left(z+b\right)}.
\ee 
The special function has the periodicity properties:
\beq
\elliptic[\displaystyle]ab{z+1}\tau=e^{i2\pi a}\elliptic[\displaystyle]abz\tau\nonumber \\
\elliptic[\displaystyle]ab{z+\tau}\tau=e^{-i\pi [\tau+2(z+b)]}\elliptic[\displaystyle]abz\tau
\eeq
and the only zero of $\elliptic[\displaystyle]abz\tau$ inside the principal region lies at $z_0=({1\over 2}-a)\tau+{1\over 2}-b$.

Recently, Haldane has proposed~\cite{Haldane18} another building block called the ``modified Weierstrass sigma function,'' whose advantage is that it only depends on the lattice $\Lambda=\left\{nL_1+mL_2|n,m\in\mathbb{Z}\right\}$,
rather than a specific $\tau$, and is thus explicitly modular covariant.
We show in Appendix~\ref{app:Sigma_functions} how to reformulate the Jain wave functions in terms of the modified Weierstrass sigma function. 

It turns out that for the analytic derivation of the Hall viscosity,
it is most useful to use the ``$\tau$ gauge'' and to continue using the Jacobi theta functions as the building blocks.
We now describe the details of the $\tau$ gauge.

The vector potential for the $\tau$ gauge is defined as 
\be
(A_x,A_y)=B\left(y,-{\tau_1 \over \tau_2}y\right),
\ee in physical coordinates, corresponding to a magnetic field $\vec{B}=-B\hat{\vec{z}}$.
The vector potential in the reduced coordinates is given by 
$(A_1,A_2)=(L_1A_x, L_1(\tau_1A_x+\tau_2A_y))=(2\pi N_\phi B\theta_2,0)$.
 In Eq.~\ref{BC}, when deforming $\tau$,
one must also move the $x$ and $y$ coordinates to account for the new geometry,
which in turn triggers a gauge transformation relating the new coordinates to the old ones.
One of the advantages of the $\tau$ gauge is that, in terms of reduced coordinates,
it is $\tau$ independent, and thus no additional gauge transformation is needed when $\tau$ is varied.
This feature is also shared by the symmetric gauge, but,
as will be demonstrated in section Sec.~\ref{sec:Analytical_proof},
the $\tau$ gauge is still more advantageous for analytic purposes.

The quasi-periodic boundary conditions are defined as
\be
\label{pbc}
t(L_i)\psi(z,\bar{z})=e^{i\phi_i}\psi(z,\bar{z}) \quad i=1,2.
\ee
The magnetic translation operator $t(\xi)$ is given by
\be
\label{magnetic translation operator}
t\left(\alpha L_1+\beta L_2\right)=e^{\alpha\partial_1+\beta\partial_2+\i2\pi\beta N_{\phi}\theta_1},
\ee
where $\partial_j\equiv{\partial \over {\partial \theta_j}}$.
The number of flux quanta through the torus is fixed to be an integer $N_\phi={V\over2\pi l^2}$,
to ensure the commutation relation $\left[t(L_1),t(L_2)\right]=0$.

We now reformulate the wave functions of Ref.~\cite{Pu17} in $\tau$ gauge.
We write the single-particle orbital in the LLL with periodic boundary conditions according to Eq.~\ref{pbc} as:
\ba
\psi^{(k)}_0(z,\bar{z})
&=&\mathcal N e^{\i\pi\tau N_\phi \theta_2^2}f^{(k)}_0(z),
\ea
\ba
\label{fsingle}
f^{(k)}_0(z)&=&\elliptic{{k\over N_\phi}+{\phi_1\over 2\pi N_\phi}}{-{\phi_2\over 2\pi}}{N_\phi z\over L_1}{N_\phi \tau},
\ea
where the subscript ``0'' refers to the LLL and the normalization factor
$\mathcal N=1/\sqrt{\ell_B L_1\sqrt{\pi}}=\left(2 N_\phi/\tau_2\right)^{1\over 4}/L_1$ is with respect to the physical coordinates.
The normalization with respect to the reduced coordinates carries an extra volume factor and is $\mathcal N=\left(2 \tau_2 N_\phi\right)^{1\over 4}$. Notably, $f_0^{(k)}$ is holomorphic in $z$.
This form of a single-orbital function has been used by L\'evay in Ref.~\cite{Levay95}.
Eq.~\ref{fsingle} has its $N_\phi$ zeros at $z_m=L_1\tau\left({1\over 2}-{k\over N_\phi}-{\phi_1\over 2\pi N_\phi}\right)+{L_1\over N_\phi}\left({1\over 2}+m+{\phi_2\over 2\pi}\right)$, with $m=0,1,\dots N_\phi-1$.
The superscript $k=0,1,2\dots N_\phi-1$ represents momentum under magnetic translations by $t(L_1/N_\phi)$,
\be
\label{T1}
t\left(L_1/N_\phi\right)\psi_0^{(k)}(z,\bar{z}) =e^{\i{{\phi_1+2\pi} k\over N_\phi}}\psi_0^{(k)}(z,\bar{z}),
\ee
while $t\left(L_2/N_\phi\right)$ changes $\psi_0^{(k)}(z,\bar{z})$ to $\psi_0^{(k+1)}(z,\bar{z})$ as
\be
\label{T2}
t\left(L_2/N_\phi\right)\psi_0^{(k)}(z,\bar{z})= e^{\i{\phi_2\over N_\phi}}\psi_0^{(k+1)}(z,\bar{z}).
\ee
From here on we will assume $\phi_{1,2}=0$ unless otherwise specified.
(We note that the single particle orbital in Eq.~\ref{fsingle} is slightly different from and more compact than the form used in Ref.~\cite{Pu17}, where it was necessary to assign zeros artificially.)

The single-particle wave function in the $n$th LL is given by
\ba
&&\psi_{n}^{(k)}(z,\bar{z})\nonumber\\
&=&\frac{\left(a^{\dagger}\right)^{n}}{\sqrt{n!}}\psi_{0}^{(k)}\nonumber\\
&=&{1\over \sqrt{n!}}\left[-\sqrt{2}\left(\ell_B\partial_z-{\bar{\tau}L_1\theta_2\over 2\ell_B}\right)\right]^n\psi_{0}^{(k)}\nonumber\\
&=&{1\over \sqrt{n!}}\mathcal N e^{\i\pi\tau N_\phi \theta_2^2}\left[-\sqrt{2}\ell_B\partial_z-{i\sqrt{2} L_1\theta_2\tau_2\over \ell_B}\right]^nf^{(k)}_0(z)\nonumber\\
&=&{1\over \sqrt{n!}}\mathcal N e^{\i\pi\tau N_\phi \theta_2^2}\sum_{t\in\mathbb{Z}+\frac{k}{N_{\phi}}}e^{\i\pi N_{\phi}\tau t^2}e^{\i2\pi N_{\phi}t\frac{z}{L_1}}\left[-{\i\over \sqrt{2}}\left(2\lambda-{\partial_\lambda}\right)\right]^n\cdot 1\nonumber\\
&=&e^{\i\pi N_{\phi}\tau \theta_2^2}f_n^{(k)}(z,\bar{z}),
\ea
where we define $\lambda\equiv {\tau_2L_1\over \ell_B}(\theta_2+t)$, and $f_n^{(k)}(z,\bar{z})$ is given by
\begin{widetext}
\be
\label{eq:psi_n_basis}
f_n^{(k)}(z,\bar{z})=
\mathcal{N}_{n}\sum_{t\in\mathbb{Z}+\frac{k}{N_{\phi}}}e^{\i\pi N_{\phi}\tau t^2}e^{\i2\pi N_{\phi}t\frac{z}{L_1}}
H_{n}\left(\frac{\tau_{2}L_1}{\ell_{B}}\left(\theta_2+t\right)\right),
\ee
with $\mathcal{N}_{n}=\frac{1}{\sqrt{2^{n}n!\ell_BL_1\sqrt{\pi}}}$ and $H_{n}(x)$ being the Hermite polynomial.
The wave function of $m$ filled LLs is just the Slater determinant of occupied single particle orbitals:
\be
\label{eq:Psi_n}
\Psi_m[z_i,\bar{z}_i]=e^{\i\pi\tau N_\phi^* \sum_{j=1}^{N}\theta_{2,j}^2}{1\over \sqrt{N}}\chi_m[f_{i}(z_j,\bar{z}_j)].
\ee
where $N_\phi^*=N/m$ is the number of flux quanta through the torus,
and $\chi_m[{f}_{i}(z_j)]$ is the determinant of $m$ filled LLs:
\be
\label{upro chi}
\chi_m[f_{i}(z_j,\bar{z}_j)]=
\begin{vmatrix}
f_0^{(0)}(z_1)&f_0^{(0)}(z_2)&\ldots&f_0^{(0)}(z_N) \\
f_0^{(1)}(z_1)&f_0^{(1)}(z_2)&\ldots &f_0^{(1)}(z_N)\\
\vdots&\vdots&\vdots \\
f_0^{(N_\phi^*-1)}(z_1)&f_0^{(N_\phi^*-1)}(z_2)&\ldots&f_0^{(N_\phi^*-1)}(z_N) \\
f_1^{(0)}(z_1,\bar{z}_1)&f_1^{(0)}(z_2,\bar{z}_2)&\ldots&f_1^{(0)}(z_N,\bar{z}_N) \\
f_1^{(1)}(z_1,\bar{z}_1)&f_1^{(1)}(z_2,\bar{z}_2)&\ldots &f_1^{(1)}(z_N,\bar{z}_N)\\
\vdots&\vdots&\vdots \\
f_{m-1}^{(N_\phi^*-1)}(z_1,\bar{z}_1)&f_{m-1}^{(N_\phi^*-1)}(z_2,\bar{z}_2)&\ldots&f_{m-1}^{(N_\phi^*-1)}(z_N,\bar{z}_N)\\
\end{vmatrix}.
\end{equation}
\end{widetext}
Here, $f_i$ represents $f_{n_i}^{(k_i)}$; i.e., the subscript $i$ of $f_i$ collectively denotes the quantum numbers $\{n_i,k_i\}$,
where $n_i$ is the LL index and $k_i$ is the momentum quantum number.  
(Note that the subscript of the many-particle wave function $\Psi_\nu$ or $\chi_n$ refers to the filling factor,
whereas the subscript of the single-particle wave function, such as $\psi^{(k)}_n$ or $f^{(k)}_n$, refers to the LL index.)
We use the convention that the LL index takes values $n=0, 1, \cdots$, with $n=0$ corresponding to the LLL,
while the filling factor has $m=1,2,\cdots$; we hope that the meaning is clear from the context.
Specifically, the LLL wave function can also be written in a Laughlin form:
\ba
\label{1LL}
\Psi_{1}[z_i,\bar{z}_i]&=&\mathcal N_1e^{\i\pi\tau N\sum_i \theta_{2,i}^2}\elliptic{{N-1\over2}}{{N-1\over2}}{Z \over L_1}{\tau}\nonumber\\
&&\quad\times\prod_{i<j}\elliptic{{\frac12}}{{\frac12}}{z_{i}-z_{j}\over L_1}{\tau},
\ea
where $Z=\sum_{i}z_{i}$, and $\mathcal N_1=\frac{\tau_{2}^{\frac{N}{4}}\eta\left(\tau\right)^{-\frac{N\left(N-3\right)}{2}-1}}{\sqrt{N!}\left(2N\pi^{2}\right)^{\frac{N}{4}}}$~\cite{Fremling16b}.
The general composite fermion wave functions before projection to LLL are written as 
\be \Psi^{\rm unproj}_{n\over 2pn+1}=\Psi_n\Psi_1^{2p},\;\Psi^{\rm unproj}_{n\over 2pn-1}=\Psi_n^{*}\Psi_1^{2p},
\ee
where we omit the normalization factors.
A particularly nice feature of working with the reduced coordinates $\theta_1$ and $\theta_2$ is that the magnetic length never enters into any expression,
other than the normalization factor.
This means that when wave functions are multiplied, no explicit rescaling of the magnetic length is needed to preserve the boundary conditions.
Displaying explicitly the exponential factors, we have
\begin{widetext}
\be
\label{CF product}
\Psi^{\rm unproj}_{n\over 2pn+1}=e^{\i\pi\tau N_\phi\sum_i \theta_{2,i}^2}\chi_n[{f}_{i}(z_j)]\left(\elliptic{{N-1\over2}}{{N-1\over2}}{Z/L_1}{\tau}\prod_{i<j}\elliptic{\frac{1}{2}}{\frac{1}{2}}{(z_{i}-z_{j})/L_1}{\tau}\right)^{2p},
\ee 
\be
\label{CF product2}
\Psi^{\rm unproj}_{n\over 2pn-1}=
e^{\left(\i\pi\tau N_\phi-2\pi|N_\phi^*|\tau_2\right)\sum_i \theta_{2,i}^2}
(\chi_n[{f}_{i}(z_j)])^{*}
\left(\elliptic{{N-1\over2}}{{N-1\over2}}{Z/L_1}{\tau}\prod_{i<j}\elliptic{\frac{1}{2}}{\frac{1}{2}}{(z_{i}-z_{j})/L_1}{\tau}\right)^{2p},
\ee
\end{widetext}
where $\chi_n[{f}_{i}(z_j)]$ is the determinant of $n$ filled LLs at the effective flux quanta $N^*_\phi=\pm N/n$ defined in Eq.~\ref{upro chi}. Note that in Eq.~\ref{CF product},
in the exponential factor the terms from different factors combine as $\tau N_\phi^*+\tau 2p N = \tau N_\phi$,
where we have used that for $\nu=1$ we have $N_\phi=N$. On the other hand,
in Eq.~\ref{CF product2}, which refers to the reverse flux states at $\nu=n/(2pn-1)$,
these terms combine as $-\bar{\tau} |N_\phi^*|+\tau 2p N = \tau N_\phi+2i \tau_2|N_\phi^*|$, with $N_\phi=2pN-|N_\phi^*|$.

The LLL projection of these wave functions requires LLL projection of products of single-particle wave functions of the type $\psi_n \psi$, where $\psi$ is some LLL wave function.
Following Refs.~\onlinecite{Pu17} and \onlinecite{Fremling19}, the LLL projection is given by
\be
\PLLL \psi_n \psi_0=\left(2\i\ell_B \over N_\phi^*\right)^{n}
  \frac{\mathcal{N}_{n}}{\mathcal{N}_{0}}e^{-\i\pi\tau N_\phi \theta_2^2} \hat{f}_n f,
\ee
where $\hat{f}_n$ now is an operator acting on $f$, which is the holomorphic part of $\psi$.
In general this operator has the form $\hat{f}_n=\hat D^{n}f_{0}$, where
\be
\label{eqn:D_operator}
\hat D=N_\phi^*\hat{\partial}_{z}-\left(N_{\phi}-N_\phi^*\right)\tilde{\partial}_{z},
\ee
has two different types of derivatives: $\tilde{\partial}_{z}$ and $\hat{\partial}_{z}$.
The first,
$\tilde{\partial}_{z}$, is understood to act only on $f_0$ and thus has the property $\tilde{\partial}_{z}f_0f=f\tilde{\partial}_zf_0$.
The second, $\hat{\partial}_{z}$, does not act on $f_0$ at all and can be defined as $\hat{\partial}_{z}f_0f=f_0\hat{\partial}_{z}f$. 
Written out explicitly, for the $n=1$ Landau level we have
\be
\hat{f}_1^{(k)}(z)\propto(N_\phi^*-N_\phi)\frac{\partial f_0^{(k)}(z)}{\partial z}+N_\phi^*f_0^{(k)}(z)\frac{\partial}{\partial z},
\ee
which has exactly the same form as Eq. 54 in Ref.~\cite{Pu17}.

It is now straightforward to apply the modified PWJ projection~\cite{Jain97,Jain97b} as shown in Ref.~\cite{Pu17}.
For the $\nu={n\over 2pn+1}$ states, the full LLL wave function is written as
 \be
 \label{projected wf}
 \Psi_{n\over 2pn+1}[z_i,\bar{z_i}]=
 e^{\i\pi\tau N_\phi\sum_i \theta_{2,i}^2}\mathcal N_1^{2p}F_1^{2p}(Z)
 \chi_n[\hat{g}_{i}(z_j)J_j^p],
\ee
\be
\label{chi-det}
{\chi_n}[\hat{g}_{i}(z_j)J_j^p]=
\begin{vmatrix}
\hat{g}_0^{(0)}(z_1)J_1^p&\ldots&\hat{g}_0^{(0)}(z_N)J_N^p \\
\vdots&\vdots&\vdots \\
\hat{g}_1^{(0)}(z_1)J_1^p&\ldots&\hat{g}_1^{(0)}(z_N)J_N^p \\
\vdots&\vdots&\vdots \\
\end{vmatrix},
\ee
where $F_1(Z)=\elliptic{{N-1\over2}}{{N-1\over2}}{Z/L_1}{\tau}$,
$J_i=\prod_{j(j\neq i)}\vartheta(z_{ij})$, and
$\vartheta(z_{ij})=\elliptic{\frac{1}{2}}{\frac{1}{2}}{(z_{i}-z_{j})/L_1}{\tau}$.
The $\hat{g}_n^{(k)}(z_i)$ is obtained from $\hat{f}_n^{(k)}(z_i)$ by making the replacement
$\partial/\partial z_i\rightarrow \redTwo\partial/\partial z_i$ for all derivatives acting on $J^p_i$.
This amounts to changing $\hat{\partial}_z\to\redTwo\hat{\partial}_z$ in Eq.~\ref{eqn:D_operator}.
For the LLL, $\hat{g}_0^{(k)}(z_i)=f_0^{(k)}(z_i)$.
For the first and second Landau levels, we explicitly have 
\be
 \label{2nd LL matrix element}
 \hat{g}_1^{(k)}(z)\propto(N_\phi^*-N_\phi)\frac{\partial f_0^{(k)}(z)}{\partial z}+N_\phi^*f_0^{(k)}(z)\redTwo\frac{\partial}{\partial z},
 \ee
\begin{widetext}
\be
\hat{g}_2^{(k)}(z)\propto (N_\phi-N_\phi^*)^2{\partial^2 f_0^{(k)}(z)\over\partial z^2}
-2N_\phi^*(N_\phi-N_\phi^*){\partial f_0^{(k)}(z)\over\partial z}\redTwo{\partial\over \partial z}
+N_\phi^{*2}f_0^{(k)}\left(\redTwo{\partial\over \partial z}\right)^2,
\ee
\end{widetext}
and projection involving yet higher LLs can be derived analogously.
We note that it is crucial for the projection that the composite fermion state be a proper state,
\ie that there be no vacant $\Lambda$-level orbitals directly underneath any occupied $\Lambda$-level orbital.
If this condition is not met, periodic boundary conditions are not necessarily preserved.

\section{Hall viscosity from microscopic Wave functions: analytical approach}
\label{sec:Analytical_proof}

The Hall viscosity is conjectured to be related to the shift in the spherical geometry.
For the Jain wave functions, the shift can be derived straightforwardly in two steps.
First one can show that the shift for the unprojected wave function $\Psi_n\Psi_1^{2p}$ is equal to the sum of the shifts of the individual factors.
From the result that the shift for $\Psi_n$ is $\sh=n$,
we obtain the result that the shift for the product is $\sh=n+2p$.
The second step is to show that the shift is preserved when the wave function is projected into the LLL. This is obviously the case,
because the LLL projection in the spherical geometry keeps the system in the same Hilbert space.

This suggests a possible route to deriving the Hall viscosity for the FQH states at $\nu=n/(2pn+1)$, following the same two steps.
We show in this section that we can accomplish the two steps provided that we assume a certain property for the normalization factor, which is justified by numerical calculation.
While the shift on the sphere is an exact property even for finite systems, the Hall viscosity of the FQH states varies with $N$,
becoming equal to the expression in Eq.~\ref{hall visc2} only in the thermodynamic limit.

\subsection{Hall viscosity for the unprojected wave functions}

The $\tau$ gauge is most favorable for an analytical proof due to the following result: 

{\em Theorem:} The normalized integer quantum Hall state $\Psi_n$, in the $\tau$ gauge, has the following property:
\be
\label{eq:int_eigen}
(\partial_{\tau_2})_\tau\Psi_n={Nn\over 4\tau_2}\Psi_n, 
\ee
where we treat, formally, $\tau$ and $\tau_2$ as the two independent variables
(which amounts to replacing $\bar{\tau} \rightarrow \tau-2i\tau_2$).

{\em Proof:} The single-particle orbital in the $\tau$ gauge,
given by Eq.~\ref{eq:psi_n_basis}, may be written as
\be
f_{n}^{(i)}=\mathcal{N}_{n}\sum_{k\in\mathbb{Z}+{i\over N_\phi}}h_{n,k}(\tau_2)\zeta_k(\tau),
\ee
where $i$ is the momentum index, $n$ is the LL index,
$\mathcal{N}_n={1\over \sqrt{2^{n}n!\sqrt{\pi}L_1\ell_B}}$,
$\zeta_k(\tau)=e^{i\pi N_\phi k(\tau k+2(\theta_1+\theta_2 \tau))}$,
and we define $h_{n,k}(\tau_2)=H_n({\tau_2L_1\over \ell_B}(\theta_2+k))$.
With some algebra, where we remember that $\tau_2L/\ell_B\propto \sqrt{\tau_2}$ and
$\partial_{\tau_{2}}\mathcal{N}_{n}=\frac{1}{4\tau_{2}}\mathcal{N}_{n}$,
together with the relation $H'_n(x)=2nH_{n-1}(x)$, we get:
\be
(\partial_{\tau_2})_\tau f_{n}^{(i)}={1+2n\over 4\tau_2}f_{n}^{(i)}
+{n(n-1)\over\tau_2}{\mathcal{N}_n\over\mathcal{N}_{n-2}}f_{n}^{(i)}.
\ee
To obtain this result, it is important to remember that the area $V=\tau_2L_1^2$ is held fixed under the variation of $\tau_2$,
such that $\partial_{\tau_{2}}\frac{\tau_{2}L_1}{\ell_{B}}\left(\theta_2+k\right)
=\partial_{\tau_{2}}\sqrt{\tau_{2}V/\ell_B^2}\left(\theta_2+k\right)
=\frac{1}{2\tau_{2}}\sqrt{\tau_{2}V/\ell_B^2}\left(\theta_2+k\right)
=\frac{L_1}{2\ell_{B}}\left(\theta_2+k\right)$.
Therefore, when $(\partial_{\tau_2})_\tau$ acts on the wave function for $n$-filled LLs $\Psi_n$, we have
\be
(\partial_{\tau_2})_\tau\Psi_n=e^{\i\pi\tau N_\phi^* \sum_i \theta_{2,i}^2}{1\over \sqrt{N}}(\partial_{\tau_2})_\tau\chi_n,
\ee
\begin{widetext}
\be
\label{un chi}
(\partial_{\tau_2})_\tau\chi_n=(\partial_{\tau_2})_\tau\mathcal{A}\prod_{j=1}^Nf_{n_j}^{(i_j)}=\mathcal{A}\sum_{j}^N\left[{1+2n_j\over 4\tau_2}f_{n_j}^{(i_j)}
  +{n_j(n_j-1)\over\tau_2}{\mathcal{N}_{n_j}\over\mathcal{N}_{n_j-2}}f_{n_j-2}^{(i_j)}\right]\prod_{j'\neq j}^Nf_{n_{j'}}^{(i_{j'})}={Nn\over 4\tau_2}\chi_n.
\ee
\end{widetext}
The last equality follows because the second term in the square brackets vanishes due to antisymmetrization. Q.E.D.

This result can be used to derive the Hall viscosity straightforwardly, as follows. 

{\em Theorem:} If a {\em normalized} wave function $\Psi$ satisfies  
\be
\label{eqngen}
(\partial_{\tau_2})_\tau\Psi={N\sh\over 4\tau_2}\Psi, 
\ee
then its Hall viscosity is given by Eq.~\ref{hall visc}. For the special case of $\Psi_n$, Eq.~\ref{eq:int_eigen} implies 
$\eta^A={\hbar Nn \over 4 V}$.

{\em Proof:} The Hall viscosity is related to the Berry curvature in Eq.~\ref{BC}, which can be expressed as
\be
\mathcal{F}_{\tau_1,\tau_2}=\partial_{\tau_1} A_2-\partial_{\tau_2} A_1,
\label{eq:Berry_connection}
\ee
where the Berry connections are defined as
\be
A_j=i\langle\Psi|\partial_{\tau_j}|\Psi\rangle,
\ee
with $\tau_1$ and $\tau_2$ chosen as independent variables. 
These can be expressed as 
\ba
\label{eq:A_to_A}
A_1&=&A_\tau+A_{\bar\tau},\nonumber\\
A_2&=&\i(A_\tau-A_{\bar\tau}).
\ea
where $A_\tau=i\langle\Psi|(\partial_{\tau})_{\bar{\tau}}|\Psi\rangle$ and
$A_{\bar\tau}=i\langle\Psi|(\partial_{\bar\tau})_\tau|\Psi\rangle$, with  
$\tau$ and $\bar\tau$ chosen as independent variables. 
With the help of Eq.~\ref{eqngen}, it is straightforward to derive the Berry connections as:
\ba
A_{\tau}&=&i\langle\Psi|\partial_\tau|\Psi\rangle
=-{1\over 2}\langle \left({\partial \over \partial \tau_2}\right)_\tau \Psi|\Psi \rangle=-{N\sh\over 8\tau_2},\nonumber\\
A_{\bar{\tau}}&=&i\langle\Psi|\partial_{\bar{\tau}}|\Psi\rangle
=-{1\over 2}\langle \Psi|\left({\partial \over \partial \tau_2}\right)_\tau\Psi \rangle=-{N\sh\over 8\tau_2}.
\ea
Eq.~\ref{eq:A_to_A} gives $A_1=-{N\sh\over 4\tau_2}$ and $A_2=0$, which implies  
\ba
\mathcal{F}_{\tau_1,\tau_2}=\partial_{\tau_1} A_2-\partial_{\tau_2} A_1=-{N\sh\over 4\tau_2^2},
\ea
finally producing $\eta^A={\hbar N\sh \over 4 V}$. Q.E.D.

We next prove the following theorem.

{\em Theorem:} The Hall viscosity for the unprojected wave function
\be
\label{eq:unproj}
\Psi_{n\over 2pn+1}^{\rm unproj}=Z\mathcal{P}_M\Psi_n\Psi_1^{2p} 
\ee
is given, in the thermodynamic limit, by 
\be
\lim_{N\rightarrow \infty}\eta^A={\hbar N (n+2p)\over 4V},
\label{etaAunproj}
\ee
provided we assume that the normalization factor $Z$ satisfies the condition
\be
\lim_{N\rightarrow \infty} {1\over N} \left({\partial\over \partial \tau_2}\right)_\tau\ln Z = 0.
\label{Z_infty}
\ee
Here, we note that $\Psi_n$ and $\Psi_1$ in Eq.~\ref{eq:unproj} are already taken as normalized,
and $Z$ is the additional factor needed for the normalization of $\Psi_{n\over 2pn+1}^{\rm unproj}$.
$\mathcal{P}_M$ projects the wave function into a definite momentum sector.

{\em Proof:}  First of all, it is a straightforward exercise to show (see Appendix~\ref{App:PLLL_vs_Mom_Proj}) that in the $\tau$ gauge we have
\be
[(\partial_{\tau_2})_\tau,\mathcal{P}_M]=0.
\label{eq:momcom}
\ee
With Eq.~\ref{eq:int_eigen} it then follows that
\begin{widetext}
\be
\label{eq:derivative_unproj}
(\partial_{\tau_2})_\tau \Psi_{n\over 2pn+1}^{\rm unproj}=   \left[\left({\partial\over \partial \tau_2}\right)_\tau\ln Z\right]  \Psi_{n\over 2pn+1}^{\rm unproj}+   Z\mathcal{P}_M\big[\big((\partial_{\tau_2})_\tau \Psi_n\big)\Psi_1^{2p}+\Psi_n\big((\partial_{\tau_2})_\tau \Psi_1^{2p}\big)\big]
\Rightarrow {(n+2p)N\over 4\tau_2}\Psi_{n\over 2pn+1}^{\rm unproj}
\ee
\end{widetext} 
where in the last step we have taken the limit $N\rightarrow\infty$ and retained the dominant term.
The Hall viscosity in Eq.~\ref{etaAunproj} follows according to our previous theorem. Q.E.D.

The derivation depends on the assumption given by Eq.~\ref{Z_infty}.
The following considerations indicate that Eq.~\ref{Z_infty} is valid: 

$\bullet$ We evaluate the $\tau$ derivative of $Z$ numerically for $\nu=2/5$ and show that it satisfies Eq.~\ref{Z_infty}; the results are shown in Appendix \ref{app:last}. 

$\bullet$ In the next section we evaluate the Hall viscosity directly for many unprojected Jain wave functions and find that they satisfy Eq.~\ref{etaAunproj} in the thermodynamic limit. 

$\bullet$ Eq.~\ref{Z_infty} is known to be true for the Laughlin states through the use of the plasma analogy \cite{Read09,Tokatly09}. However,
an analogous plasma analogy for Jain states is likely to be much more complicated and remains an open question.

$\bullet$ In Appendix~\ref{app:last} we explicitly evaluate the contribution of the $\left({\partial\over \partial \tau_2}\right)_\tau\ln Z$  term in Eq.~\ref{eq:derivative_unproj} to the Hall viscosity.
We find that while it provides a correction for finite $N$, the correction vanishes with increasing $N$.

A comment regarding an interplay between Hall viscosity and momentum projection is in order.
The commutator $[(\partial_{\tau_2})_\tau,\mathcal{P}_M]=0$ does not guarantee that the wave functions with and without momentum projection have the same Hall viscosity, since the normalization factors can be different.
In fact, for finite systems, they do have different Hall viscosities,
since the decomposition into the various momentum states is itself $\tau_2$ dependent (see \eg Appendix~\ref{App:MomProj}).
This $\tau_2$ dependence may introduce extra Berry phases and thus alter the Hall viscosity.
In Appendix~\ref{App:Momentum_Mixing} we however show that this change in the viscosity is sub-leading in $N$ and does not therefore contribute in the thermodynamic limit.

We finally note that the above considerations apply to wave functions obtained from the parton construction~\cite{Jain89b}, which are products of integer quantum Hall wave functions.
We expect that the Hall viscosity of the wave function $\Psi_{\nu}=\prod_{\lambda=1}^m\Psi_{n_\lambda}$, with $\nu^{-1}=\sum_{\lambda=1}^m n^{-1}_{\lambda}$,
is given by Eq.~\ref{hall visc} with $\sh=\sum_{\lambda=1}^m n_\lambda$.

We note that the above proof only applies to product states where each factor has magnetic field pointing in the same direction.
Generalization to reverse-flux attached states at $\nu=n/(2pn-1)$, or to states of parton theory with negative values of $n_\lambda$, remains an open problem, because Eq.~\ref{eqngen} is not satisfied for $\Psi_n^{*}$. 

%Eq.~\ref{eqngen} requires $\Psi_n$ and $\Psi_n^{*}$ to be eigenstates of the same operater, but neither $\left(\partial_{\tau_2}\right)_\tau$ nor $\left(\partial_{\tau_2}\right)_{\bar{\tau}}$ satisfies this condition.

\subsection{Hall viscosity for the LLL projected wave functions}
Next we project the wave functions to the LLL. In this case we have the following theorem

{\em Theorem:} 
The Hall viscosity of the normalized projected wave function 
\be
\Psi_{n\over 2pn+1}=Z^\prime\PLLL\mathcal{P}_M\Psi_n\Psi_1^{2p},
\label{projwf'}
\ee
is given by Eq.~\ref{etaAunproj}, provided the normalization factor 
$Z'$ ($Z^\prime\neq Z$) satisfies the condition 
\be
\lim_{N\rightarrow \infty} {1\over N} \left({\partial\over \partial \tau_2}\right)_\tau\ln Z' = 0.
\label{Zp_infty}
\ee
Again, $\Psi_n$ and $\Psi_1$ are taken as normalized, and $\PLLL$ denotes LLL projection.

{\em Proof:} As explained in Sec.~\ref{tau wf}, the ``direct" projection is accomplished by the replacement $\chi_n\equiv\chi_n\big(f_n^{(i)}\big)\to \hat{\chi}_n\equiv\chi_n\big(\hat{D}^nf_0^{(i)}\big)$ in the unprojected wave function. We now prove:
\be
\label{LLL chi}
[\big(\partial_{\tau_2}\big)_\tau , \hat{\chi}_n]={Nn\over 4\tau_2}\hat{\chi}_n.
\ee
With this result, the Hall viscosity of the LLL-projected wave functions in Eq.~\ref{projwf'} can be obtained in the same fashion as for the unprojected wave function.

Let us first consider $(\partial_{\tau_2})_\tau \hat{D}^nf_0^{(i)}$.
We first note that $\left[\left(\partial_{\tau_2}\right)_\tau,\partial_z\right]={1\over \tau_2}\big({1\over 2}\partial_z+\partial_{\bar{z}}\big)$
since $z$ depends on $\tau_1$ and $\tau_2$.
We then get:
\be
\label{com D}
\big(\partial_{\tau_2}\big)_\tau\hat{D}=\hat{D}\big(\partial_{\tau_2}\big)_\tau+{1\over 2\tau_2}\hat{D}+{1\over \tau_2}\big[N_\phi^*\hat{\partial}_{\bar{z}}-(N_\phi-N_\phi^*)\tilde{\partial}_{\bar{z}}\big].
\ee
As $\hat{D}$ only acts on LLL wave functions, which consequentially are analytical in $z$,
the last term on the right-hand side of Eq.~\ref{com D} may be omitted.
Thereby, we effectively have:
\be
\label{eq56}
(\partial_{\tau_2})_\tau \hat{D}^nf_0^{(i)}={1+2n\over 4\tau_2}\hat{D}^nf_0^{(i)}+\hat{D}^nf_0^{(i)}(\partial_{\tau_2})_\tau.
\ee
Eq.~\ref{LLL chi} then follows because:
\ba
\big(\partial_{\tau_2}\big)_\tau\hat{\chi}_n&=&\left(\partial_{\tau_2}\right)_\tau \chi_n\big(\hat{D}^{n_j}f_0^{(i_j)}\big)
=(\partial_{\tau_2})_\tau\mathcal{A}\prod_{j=1}^N\hat{D}^{n_j}f_{0}^{(i_j)}\nonumber\\
&=&\mathcal{A}\Big\{\sum_{j}^N{1+2n_j\over 4\tau_2}\prod_{j'}^N\hat{D}^{n_{j'}}f_{0}^{(i_{j'})}\nonumber\\
&&\quad+\prod_{j'}^N\hat{D}^{n_{j'}}f_{0}^{(i_{j'})}\big(\partial_{\tau_2}\big)_\tau\Big\}
\nonumber\\
&=&{Nn\over 4\tau_2}\hat{\chi}_n+\hat{\chi}_n\big(\partial_{\tau_2}\big)_\tau.
\ea
Assuming Eq.~\ref{Zp_infty}, we now have, in the limit $N\rightarrow \infty$, 
\ba
\label{LLL visc}
(\partial_{\tau_2})_\tau \Psi_{n\over 2pn+1}&=&Z^\prime\mathcal{P}_Me^{\i\pi\tau N_\phi\sum_i \theta_{2,i}^2} (\partial_{\tau_2})_\tau\big\{\hat{\chi}_n\chi_1^{2p}\big\}\nonumber\\
&=&Z^\prime\mathcal{P}_Me^{\i\pi\tau N_\phi\sum_i \theta_{2,i}^2}\big\{{Nn\over 4\tau_2}\hat{\chi}_n\chi_1^{2p}\nonumber\\
&&+2p\hat{\chi}_n\chi_1^{2p-1}\big[(\partial_{\tau_2})_\tau\chi_1\big]\big\}\nonumber\\
&=&{(n+2p)N\over 4\tau_2}\Psi_{n\over 2pn+1}.
\ea
Q.E.D.

The above proof also proceeds analogously for the PWJ projected wave functions.
In the PWJ projection in Eq.~\ref{projected wf}, $\chi_1$ is decomposed into two parts:
the center-of-mass part $F_1(Z)$ (which includes the normalization factor $\mathcal N_1$) and the Jastrow factors.
Noting that $\left(\partial_{\tau_{2}}\right)_{\tau}\chi_1={N\over 4\tau_2}\chi_1$ and the Jastrow factors $J_i$s are analytical functions of $\tau$,
the center-of-mass part must be an eigenfunction of $\left(\partial_{\tau_{2}}\right)_{\tau}$ with eigenvalue ${N\over 4\tau_2}$.
The Jastrow factors can be incorporated into $\hat{\chi}_n$ with the change $\hat{D}\rightarrow 2N_\phi^*\hat{\partial}_{z}-\left(N_{\phi}-N_\phi^*\right)\tilde{\partial}_{z}$.
Since, {\em mutatis mutandis}, Eq.~\ref{eq56}  still holds for the new $\hat{D}$,
we have $\left(\partial_{\tau_{2}}\right)_{\tau}\chi_n\left[\hat{g}_{i}(z_j)J_j^p\right]={Nn\over 4\tau_2}\chi_n\left[\hat{g}_{i}(z_j)J_j^p\right]$.
As a final result, Eq.~\ref{LLL visc} is still valid after PWJ projection.

The proof relies on the assumption in Eq.~\ref{Zp_infty}. For the LLL projected wave functions it is nontrivial to calculate $Z'$ numerically. However,
we evaluate in the next section the Hall viscosity for the projected wave functions for several fractions and find that it converges to Eq.~~\ref{etaAunproj} in the thermodynamic limit, which suggests that Eq.~\ref{Zp_infty} is valid.
Although the equivalence of the Hall viscosity for the projected and unprojected CF states is nontrivial, the result should not come as a complete surprise. 
After all, the two states still describe the same topological phase. Our results actually support the notion that the Hall viscosity is a topological quantity.

\section{Numerical evaluations of the Hall viscosity}
\label{numerical}

We now use the Monte Carlo method to numerically evaluate the Hall viscosity defined in Eq.~\ref{berry-curv} for the wave functions constructed in Sec.~\ref{sec:wfns}.
To calculate the inner product $\big\langle {\partial \Psi \over \partial \tau_1}\big|{\partial \Psi \over \partial \tau_2}\big\rangle$,
we make the approximation
${\partial \Psi \over \partial \tau_1}\approx {\Psi(\tau_1+\delta,\tau_2)-\Psi(\tau_1-\delta,\tau_2)\over 2\delta}$ and
${\partial \Psi \over \partial \tau_2}\approx {\Psi(\tau_1,\tau_2+\tau_2\delta)-\Psi(\tau_1,\tau_2-\tau_2\delta)\over 2{\tau_2\delta}}$, with $\delta\sim 10^{-3}$. 
The function $|\Psi(\tau_1,\tau_2)|^2$ is chosen as the weight function for the calculation of all inner products.
We also make use of the ``lattice Monte Carlo" trick introduced by Wang {\em et al.}~\cite{Wang19}.

\begin{figure}[tb]
  \includegraphics[width=\columnwidth]{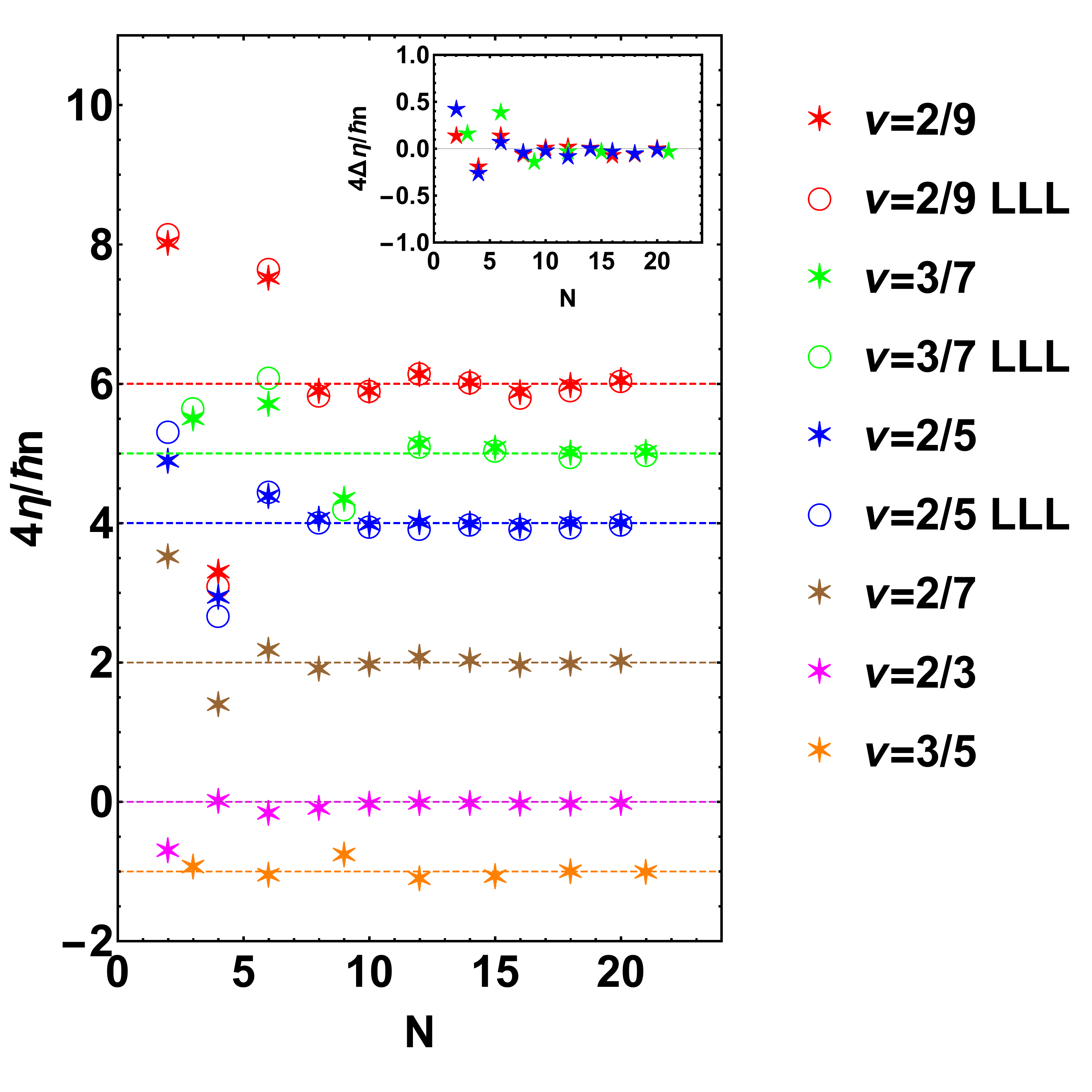} 
  \caption{Hall viscosities as a function of the system size at fillings $2/5$, $3/7$, $2/9$, $2/3$, $3/5$ and $2/7$.
The calculations are performed for a square torus with $\tau=i$.
    For $\nu={n\over 2pn+1}$,  the Hall viscosities are calculated for both the LLL-projected and unprojected wave functions, given in Eq.~ \ref{projected wf} and Eq.~\ref{CF product},
depicted by circles and stars, respectively.  For $\nu={n\over 2pn-1}$,
the Hall viscosities are calculated only for the unprojected wave functions given in Eq.~\ref{CF product2}. 
The expected quantized Hall viscosities are given by ${\hbar n\sh\over 4}$, marked by dashed horizontal lines,
where $\sh$ is the shift, and $n={N\over V}$. The Hall viscosities calculated from both the projected and the unprojected wave functions approach the expected value with increasing $N$, the number of particles.
The inset shows the difference between the Hall viscosities of the projected and the unprojected wave functions; the difference vanishes with increasing $N$.
  }
  \label{Fig1}
\end{figure}

In Fig.~\ref{Fig1} we show the Hall viscosities for various fillings calculated using Eq.~\ref{berry-curv}.
We use both LLL-projected and -unprojected wave functions to evaluate Hall viscosities at fillings $2/5$,
$3/7$ and $2/9$ for the states with positive flux attachment.
For fillings $2/3$, $3/5$ and $2/7$,
we only use the unprojected wave functions to evaluate Hall viscosities,
since the PWJ projection for reverse flux sates on the torus is much more cumbersome to implement.
There are many noteworthy conclusions that can be drawn from our study.

\begin{figure}[tb]
	\includegraphics[width=\columnwidth]{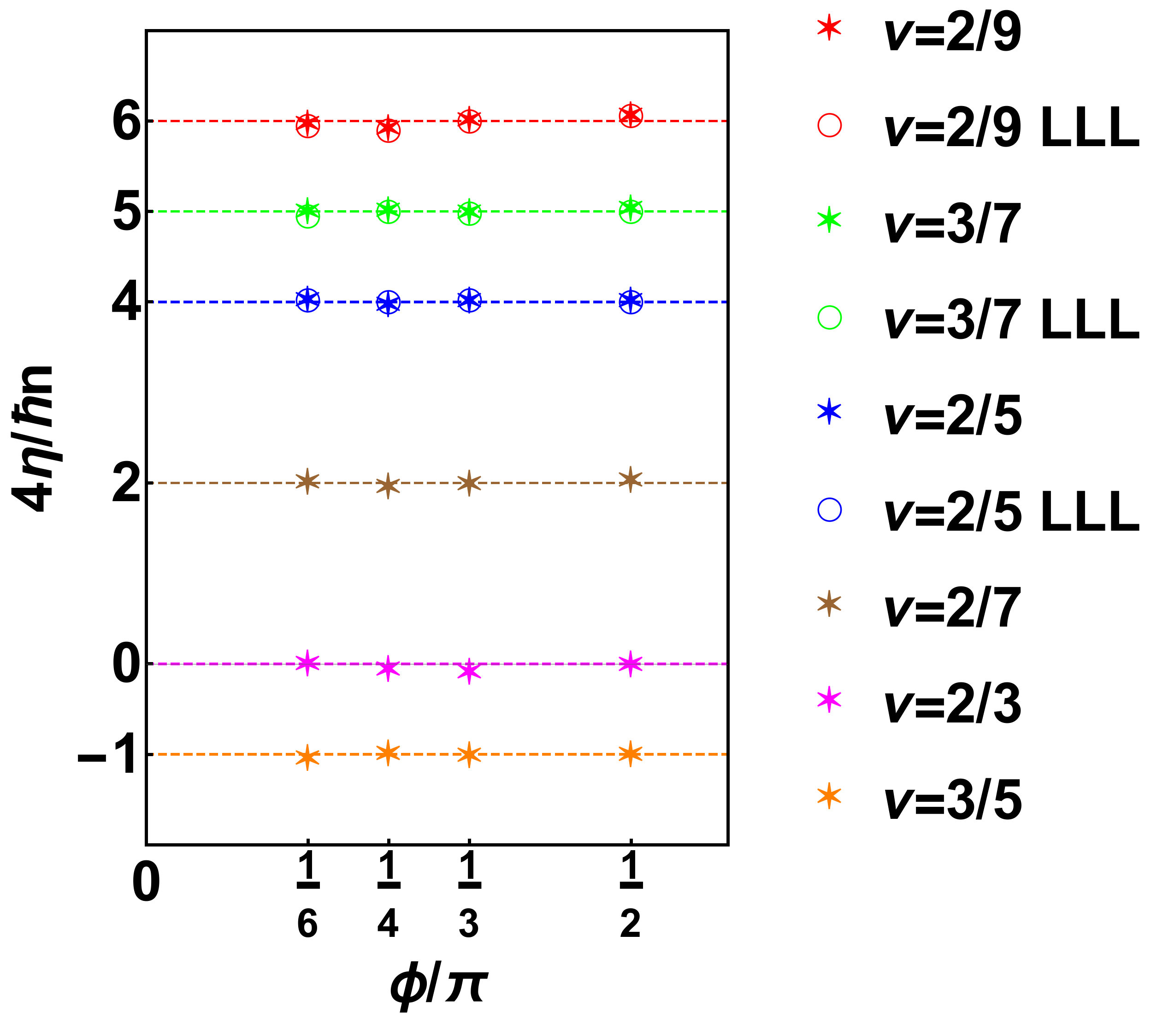} 
	\caption{Hall viscosities for $\tau=e^{i\phi}$ along the unit circle.
We choose $N=20$ for $2/5$, $2/9$, $2/3$ and $2/7$ while $N=21$ for $3/7$ and $3/5$.
The dashed lines represent the quantized values given by Eq.~\ref{hall visc}.
        }
	\label{Fig2}
\end{figure}

\begin{figure}[t]
	\includegraphics[width=\columnwidth]{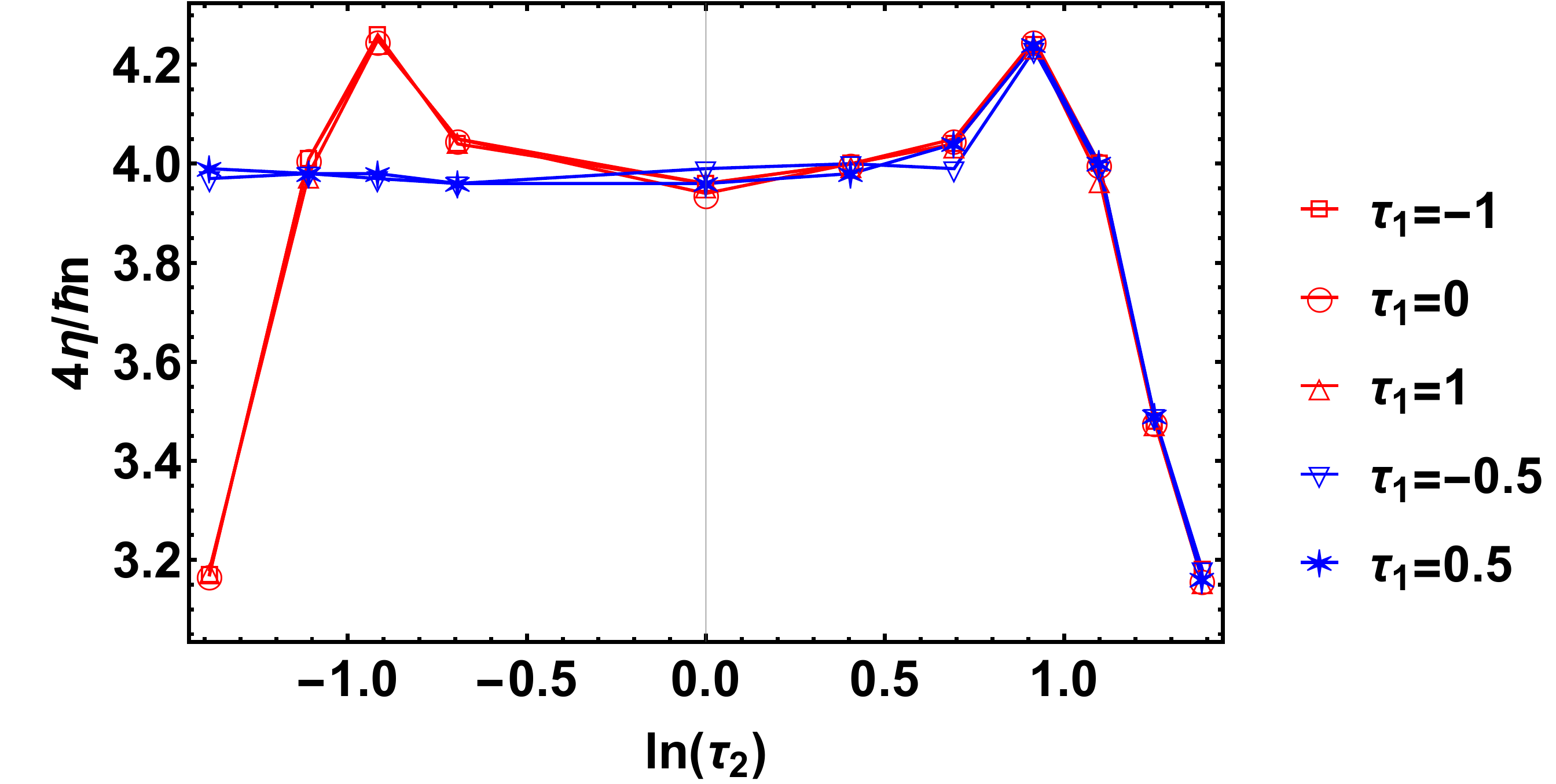} 
	\caption{
 The Hall viscosity for the 16-particle $\nu=2/5$ state at several positions in the $\tau$ plane.
 The different symbols represent different values of $\tau_1$,
 whereas the horizontal axis plots $\ln(\tau_2)$.
For a fairly large region around $\tau=\i$ the Hall viscosity is seen to be quantized at the value given by Eq.~\ref{hall visc} with $\sh=4$.
This figure illustrates that the results are consistent with the modular covariance of our wave functions.
First, Hall viscosities for points related by $\tau_1\to \tau_1+1$ coincide; there are two groups of such points: 
$\tau_1=-1,0,1$ plotted in red while $\tau_1=-0.5, 0.5$ plotted in blue. Furthermore,
the $\tau_1=0$ curve is symmetric between $-\ln(\tau_2)$ and $\ln(\tau_2)$,
consistent with the symmetry under $\tau\rightarrow -1/\tau$.
        }
	\label{Fig3}
\end{figure}

\begin{figure}
  \includegraphics[width=\columnwidth]{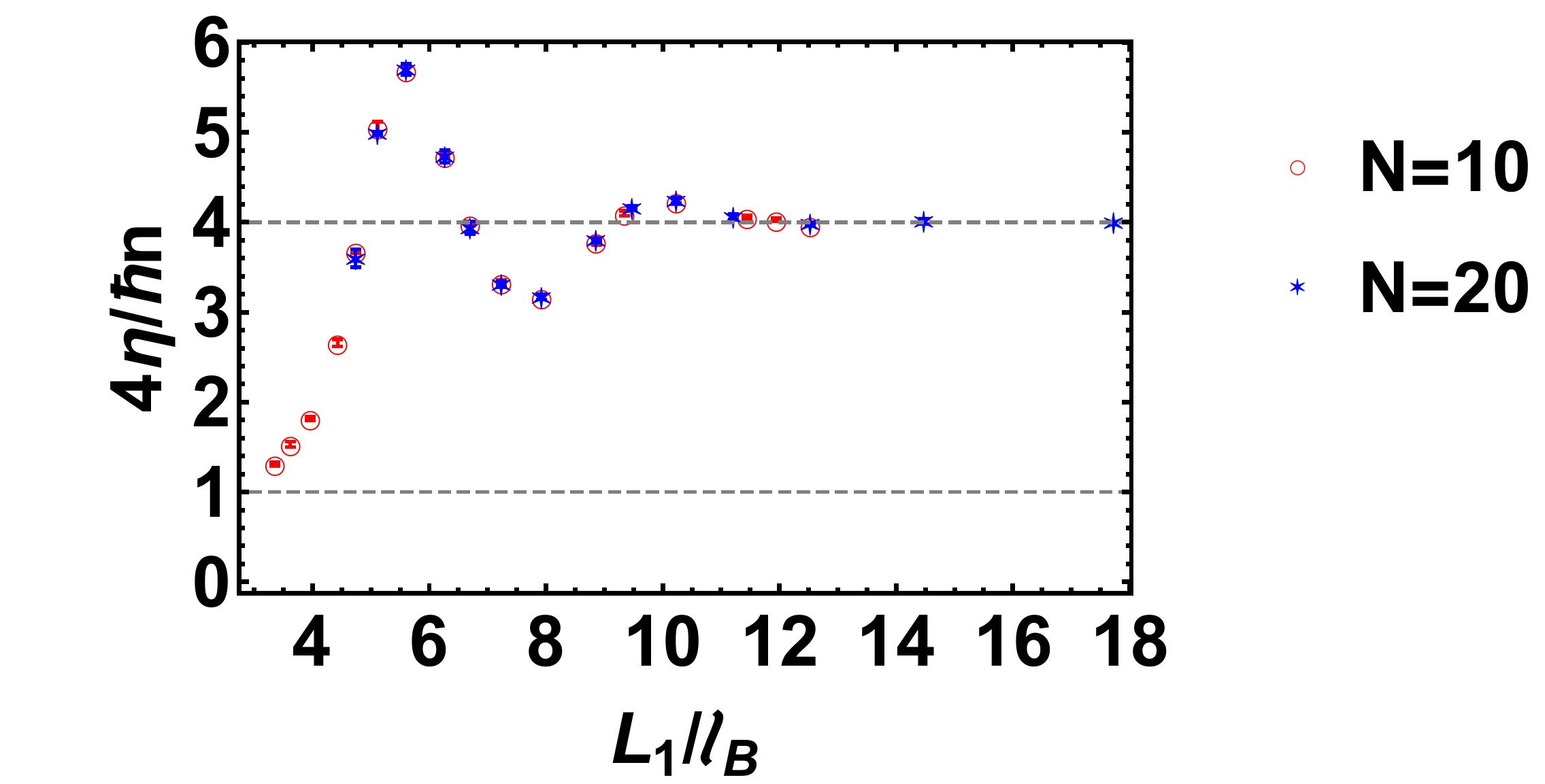} 
  \caption{Hall viscosity of the $\nu=2/5$ state as a function of $L_1$,
the shorter side of the torus. Results are given for two systems with $N=10$ and $N=20$,
for a rectangle torus with $\tau=i{V\over L_1^2}$, where $V$ is the area of torus.
The length $L_1$ is quoted in units of $l_B$. When $L_1$ becomes comparable to the magnetic length $l_B$,
there are obvious fluctuations in the Hall viscosity. In the thin-torus limit $L_1/l_B\to 0$,
the Hall viscosity approaches the value consistent with $\sh=1$.
  }
  \label{Fig4}
\end{figure}

$\bullet$ The most important conclusion from our study is that the Hall viscosities converge to the expected value in Eq.~\ref{hall visc} for sufficiently large systems.
We find that the convergence is achieved already in systems that contain on the order of ten composite fermions. Furthermore,
the Hall viscosities have the expected values for both the projected and the unprojected wave functions,
supporting the notion that they are a topological property of the state, independent of microscopic details.

$\bullet$ The Hall viscosities have significant finite-size corrections for small $N$.
As shown in the previous section,
a quantized Hall viscosity would be obtained for the unprojected wave functions if the overall normalization factor had satisfied Eq.~\ref{Z_infty} for arbitrary $N$.
Our numerical results show that this equation is satisfied only in the thermodynamic limit. 

$\bullet$ The Hall viscosities of the LLL projected and unprojected wave functions are surprisingly close (though not exactly equal),
indicating that the LLL projection does not significantly alter the Hall viscosity even for small $N$.
The small deviation between them vanishes as the system size increases, as shown in the inset of Fig.~\ref{Fig1}.

$\bullet$ The convergence of the viscosity is achieved faster for $\nu=2/5$ than for $3/7$ and $2/9$.
We attribute this to a longer correlation length for the latter two, as measured,
for example from the size of the quasiparticle or the quasihole excitations. 
As the correlation length increases, the system needs to be larger in order for a particle to ``forget'' that it resides on a finite geometry.
Indeed, a comparison with earlier works~\cite{Read10,Fremling16b} shows that the Hall viscosity for $1/3$ converges even faster than that for $2/5$.

$\bullet$  For all of the above calculations we have assumed a square torus.
We have also studied the dependence on the corner angle  as well as the aspect ratio. 
Fig.~\ref{Fig2} shows the Hall viscosities for different values of $\tau$ along the unit circle at $\tau=e^{\i\phi}$ for $N=21$ particles.
The variations with $\phi$ are very small, showing again that the Hall viscosity of the projected and unprojected wave functions is independent of the geometry of the torus provided that the system is large enough.

$\bullet$  To further test its robustness,
we scan the Hall viscosity in the $\tau$ plane for the 16 particle system at $\nu=2/5$. The result is 
 shown in Fig.~\ref{Fig3}. The Hall viscosities at different $\tau_1$ are represented by different labels,
and the horizontal axis represents $\ln(\tau_2)$.
The Hall viscosities are given by Eq.~\ref{hall visc2} for a fairly large region around $\tau=\i$,
but begins to show corrections when $\tau$ deviates too far from $\tau=\i$.
As expected from the modular covariance of the wave functions, 
the points connected by modular transformations $\tau\to \tau+1$ (whose symbols have the same color) give exactly the same Hall viscosities. 
The curve at $\tau_1=0$ is symmetric under the transformation $\tau_2\to{1\over \tau_2}$, as also expected from modular covariance.

$\bullet$ In Fig.~\ref{Fig4}, we show the computed the Hall viscosity in the thin-torus limit $\tau_2>>1$.
We consider the 2/5 Jain state for two systems with $N=10$ and $N=20$ particles.
We plot the Hall viscosity as a function of $L_1$ rather than $\tau_2$ to highlight the fact that 
the $L_1$-dependence of the viscosity is independent of the number of particles.
This independence has been noted earlier in Ref.~\onlinecite{Fremling16b} for the Laughlin state,
and also coincides with earlier work on the cylinder geometry \cite{Zaletel13}.
One can understand this size independence by considering that the Hall viscosity should be a local characteristic of the Hall fluid. As such,
whether one places the fluid on an infinite cylinder or a torus makes little difference, provided that the torus is long enough.
Thus the fluid will only be sensitive to the shorter circumference $L_1$ (the shorter edge) of the torus or cylinder,
and deviations of the viscosity are expected to occur only when $L_1$ is comparable to
or smaller than the correlation length of the fluid.
It is also noteworthy that in the thin-torus limit,
in which the composite fermions form a charge-density wave,
the Hall viscosity is seen to approach $\sh=1$.
This is expected because (a) in the thin-torus limit,
the 2/5 wave function reduces to a single Slater determinant,
and (b) all LLL slater determinants have the same viscosity, $ \eta^A={\hbar \over 4}{N \over V}$.
This is true of all FQH states that reduce to a single Slater determinant in the thin-torus limit.

\begin{figure}
	\includegraphics[width=\columnwidth]{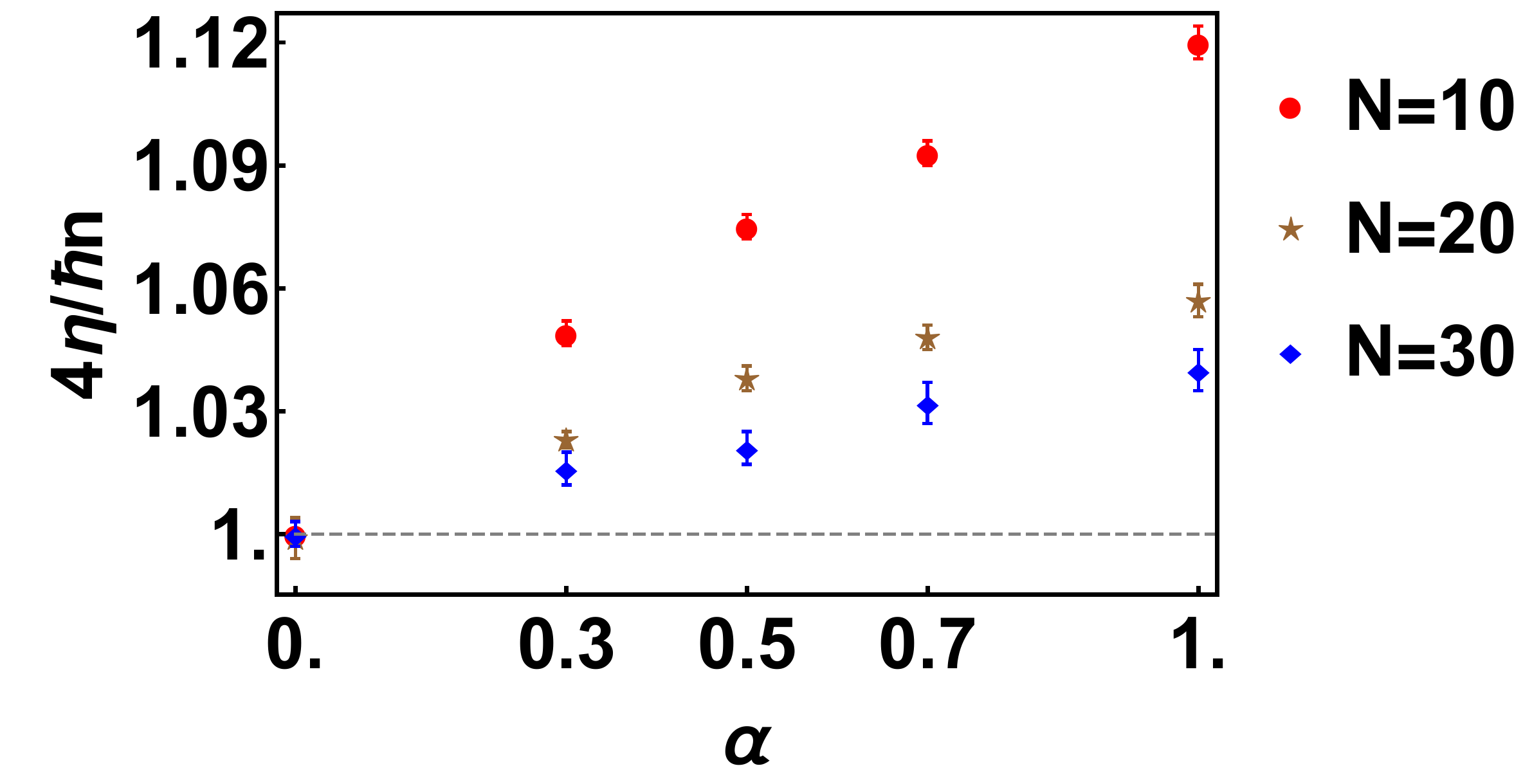} 
	\caption{
          The Hall viscosity of $\Psi_1|\Psi_1|^\alpha$ with varying $\alpha$ for different system sizes.
          The Hall viscosity for nonzero $\alpha$ appears, by visual inspection,
to approach $ \eta^A={\hbar \over 4}{N \over V}$ with increasing system size.}
	\label{Fig5}
\end{figure}

$\bullet$ We have noted above that our results are likely also applicable to the wave functions in the parton construction, which are products of integer quantum Hall states.
It is possible to further generalize the class of wave functions that produce the quantized Hall viscosity.
We consider the wave function $\Psi_1 |\Psi_1|^{\alpha}$ as a trial wave function for $\nu=1$.
This wave function has the same topological structure as $\Psi_1$, occurs at the same shift in the spherical geometry,
and may represent the physics of LL mixing due to interactions.
We find that the Hall viscosity approaches the value $ \eta^A={\hbar \over 4}{N \over V}$
in the thermodynamic limit independently of the value of $\alpha$, as shown in Fig.~\ref{Fig5}.

\section{conclusion}

In this work, we have evaluated the Hall viscosities of the Jain states at many filling factors of the form $\nu=n/(2pn\pm 1)$,
specifically for $\nu=2/5$, $3/7$, $2/9$, $2/3$, $3/5$ and $2/7$.
For this purpose we use LLL wave functions constructed in Ref.~\cite{Pu17}, but appropriately reexpressed in $\tau$ gauge.
The numerical results agree with Eq.~\ref{hall visc} proposed by Read~\cite{Read09},
which relates the Hall viscosity to the orbital spin or the shift in the spherical geometry.
We also find that the Hall viscosities for the unprojected and projected wave functions are the same in the thermodynamic limit,
supporting the notion that the Hall viscosity is a topological property of a FQH state. 

Additionally, the product form of our wave functions suggests a possible analytical derivation for the Hall viscosity of a large class of FQH states,
analogous to the derivation of the shift in the spherical geometry.
We show that the Hall viscosity of the unprojected and projected wave functions can be derived analytically provided we make an assumption regarding the thermodynamic behavior of an overall normalization factor $Z$.
This assumption has been confirmed by detailed computer calculations.

\begin{acknowledgments} The work at Penn State (S.P. and
J.K.J.) was supported in part by the U. S. Department of Energy,
Office of Basic Energy Sciences, under Grant No. DE-SC0005042.
The work at Utrecht (M.F.) is part of the D-ITP consortium,
a program of the Netherlands Organisation for Scientific Research (NWO) that is funded by the Dutch Ministry of Education, Culture and Science (OCW).
The numerical calculations were performed using Advanced CyberInfrastructure computational resources provided by the Institute for CyberScience at The Pennsylvania State University.
We are grateful to F. D. M. Haldane, P\'eter L\'evay, Johnathan Schirmer, and D. T. Son for their help and advice.
\end{acknowledgments}

\begin{appendix}

\section{Proof of the modular invariance of Hall viscosity for CF wave functions}
\label{app:mod_vis}
In this appendix, we prove that the Hall viscosity defined in Eq.~\ref{berry-curv} is modular invariant provided that Eq.~\ref{KMM} is satisfied.
The proof for the modular transformation $\tau\rightarrow \tau+1$ is trivial, since the wave function is invariant \cite{Fremling19}.
Below we show the statement is also true for $\tau\rightarrow -{1\over \tau}$. 

In Ref.~\cite{Fremling19} it is shown that under this transformation,
Eq.~\ref{KMM} holds with $K_{M,M'}={1\over \sqrt{2pn\pm 1}}e^{i2\pi{nMM'\over 2pn\pm 1}}$.
We then note that Hall viscosity is independent of the center-of-mass momentum, since:
\ba
&&\big\langle{\partial \Psi^{(M+1)}\over \partial \tau_1}\big|{\partial \Psi^{(M+1)}\over \partial \tau_2}\big\rangle\nonumber \\
&=&\big\langle{\partial t_{CM}({L_2\over N_\phi})^\dagger \Psi^{(M)}\over \partial \tau_1}\big|{\partial t_{CM}({L_2\over N_\phi})\Psi^{(M)}\over \partial \tau_2}\big\rangle\nonumber \\
&=&\big\langle{\partial \Psi^{(M)}\over \partial \tau_1}\big| t_{CM}({L_2\over N_\phi})^\dagger t_{CM}({L_2\over N_\phi})\big|{\partial \Psi^{(M)}\over \partial \tau_2}\big\rangle\nonumber \\
&=&\big\langle{\partial \Psi^{(M)}\over \partial \tau_1}\big|{\partial \Psi^{(M)}\over \partial \tau_2}\big\rangle
\ea
From the first line to second line, we used the factor that $t_{CM}({L_2\over N_\phi})$ is independent of $\tau_1$ and $\tau_2$, as shown in Appendix~\ref{App:PLLL_vs_Mom_Proj}. Similarly,
because $t_{CM}({L_1\over N_\phi})$ is also independent of $\tau_1$ and $\tau_2$, it is straightforward to see $\big\langle{\partial \Psi^{(M')}\over \partial \tau_1}\big|{\partial \Psi^{(M)}\over \partial \tau_2}\big\rangle=\delta_{M,M'}\big\langle{\partial \Psi^{(M)}\over \partial \tau_1}\big|{\partial \Psi^{(M)}\over \partial \tau_2}\big\rangle$.

Under the transformation $\tau\rightarrow -{1\over \tau}$,
we have $\tau_1'=-{\tau_1\over |\tau|^2}$ and $\tau_2'={\tau_2\over |\tau|^2}$.
The Hall viscosity in Eq.~\ref{berry-curv} thus transforms like
\ba
\eta^A&\rightarrow& {2\hbar \tau_2'^2\over V}\sum_{M'} |K_{M,M'}|^2{\rm Im}\big\langle{\partial \Psi^{(M)}\over \partial \tau'_1}\big|{\partial \Psi^{(M)}\over \partial \tau'_2}\big\rangle\nonumber \\
&=&{2\hbar \tau_2'^2\over V}{\rm Im}\big\langle{\partial \Psi^{(M)}\over \partial \tau'_1}\big|{\partial \Psi^{(M)}\over \partial \tau'_2}\big\rangle.
\ea
Given that
\beq
{\partial \Psi^{(M)}\over \partial\tau'_1}={\tau'^2_1-\tau'^2_2\over (\tau'^2_1+\tau'^2_2)^2}{\partial \Psi^{(M)}\over \partial\tau_1}-{2\tau'_1\tau'_2\over (\tau'^2_1+\tau'^2_2)^2}{\partial \Psi^{(M)}\over \partial\tau_2}\nonumber\\
{\partial \Psi^{(M)}\over \partial\tau'_2}={\tau'^2_1-\tau'^2_2\over (\tau'^2_1+\tau'^2_2)^2}{\partial \Psi^{(M)}\over \partial\tau_2}+{2\tau'_1\tau'_2\over (\tau'^2_1+\tau'^2_2)^2}{\partial \Psi^{(M)}\over \partial\tau_1},\nonumber\\
\eeq
it follows that
\ba
\eta^A&\rightarrow& {2\hbar \tau_2'^2\over V}{1\over (\tau'^2_1+\tau'^2_2)^2}{\rm Im}\big\langle{\partial \Psi^{(M)}\over \partial \tau_1}\big|{\partial \Psi^{(M)}\over \partial \tau_2}\big\rangle\nonumber \\
&=& {2\hbar \tau_2'^2\over V}{\tau_2^2\over \tau'^2_2}{\rm Im}\big\langle{\partial \Psi^{(M)}\over \partial \tau_1}\big|{\partial \Psi^{(M)}\over \partial \tau_2}\big\rangle\nonumber \\
&=&\eta^A.
\ea
This shows explicitly that the Hall viscosity is modular invariant for CF wave functions.

\begin{figure*}[t]
  \includegraphics[width=1.0\linewidth]{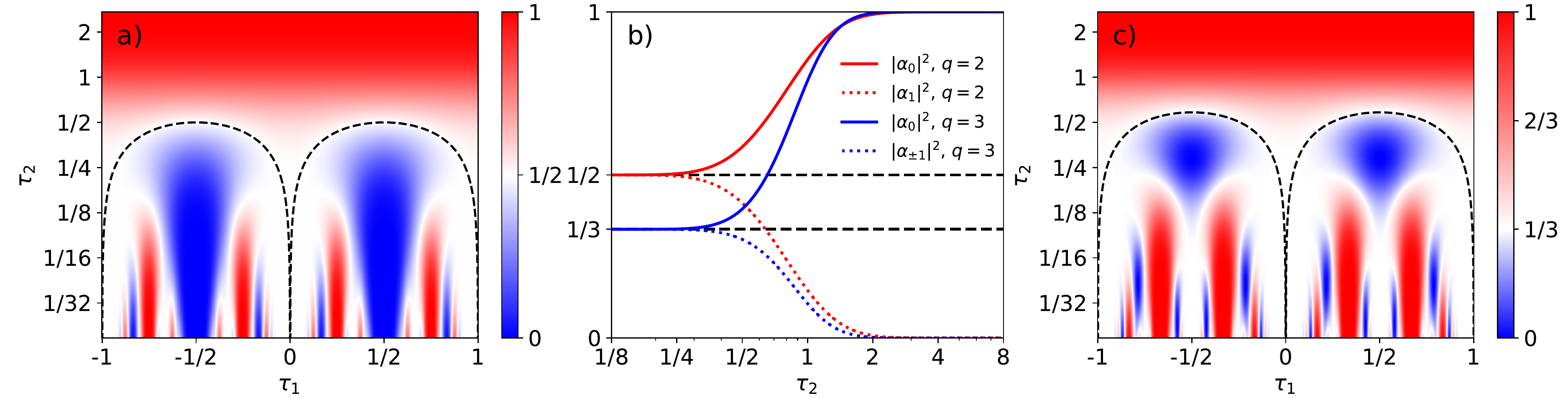}
  \caption{The absolute values of $\alpha_0^2$ for the CF Laughlin state in the whole $\tau$ plane for a) $q=2$ and c) $q=3$.
    The middle panel b) shows the components $|\alpha_k|^2$ along the line $\tau=\i\tau_2$.
    As a guide to the eye,
    panel a) and c) contains a dashed line satisfying the equation $\tau_2^2+(\tau_1\pm1/2)^3=1/4$ and $(6\tau_2/7)^2+(\tau_1\pm1/2)^2=1/4$, respectively.
  }\label{Fig6}
\end{figure*}

\section{The momentum components of the CF state}
\label{App:MomProj}

It is noted in the main text that $\Psi_{n\over 2pn+1}$ is not a momentum eigenstate,
but a superposition of $2pn+1$ different momentum eigenstates,
\ie $\Psi_{n\over 2pn+1}=\sum_{m=0}^{2p}\alpha_m\Psi_{n\over 2pn+1}^{(m)}$,
in which $\Psi_{n\over 2pn+1}^{(m)}$ are momentum eigenstates.
It turns out that the coefficients $\alpha_m$ have a dependence on both $\tau$ and $\tau_2$, which, in turn,
gives rise to an extra Berry curvature.
In this appendix and the next, using the CF formulation of the Laughlin state as an explicit example,
we show how this extra Berry curvature comes about and also that it is sub-leading in the thermodynamic limit.

We take as a starting point the two formulations of the Laughlin state at $\nu=1/q$.
The first is the momentum projected state
\ba
\label{eq:Laughlin}
&&\Psi_{1/q}^{\left(p\right)}\left(z\right)=
\mathcal{N}\left(\tau\right)e^{\i\pi\tau N_{\phi}\sum_{i}y_{i}^{2}}\nonumber\\
&&\quad\quad\quad\times\prod_{i<j}\elliptic {1/2}{1/2}{\frac{z_{ij}}{L_1}}{\tau}^{q}{F}_{1/q}^{(p)}\left(\frac{Z}{L_1},\tau\right),
\ea
 where the center of mass is 
 \[
{F}_{1/q}^{(p)}\left(Z,\tau\right)=\elliptic{\frac{p}{q}+q\frac{N-1}{2}}{q\frac{N-1}{2}}{qZ}{q\tau},
\]
and the supposed normalization is 
$\mathcal{N}\left(\tau\right)\propto\tau_{2}^{\frac{qN}{4}}$.
The second version is obtained by multiplying $q$ copies of the $\nu=1$ wave functions,
$\Phi_{1/q}\left(z\right)=\left[\Psi_{1}\left(z\right)\right]^{q}$.
The quotient between these two functions is 
\be
\frac{\Psi_{1/q}^{\left(p\right)}\left(z\right)}{\Phi_{1/q}\left(z\right)} \propto
\frac{\elliptic{\frac{p}{q}+q\frac{N-1}{2}}{q\frac{N-1}{2}}{qZ}{q\tau}}
     {\left[\elliptic {\frac{N-1}{2}}{\frac{N-1}{2}}Z{\tau}\right]^{q}}.
\ee
Thus, the only difference lies in the center of mass representations.
For simplicity we will assume that $N$ is odd such that $(N-1)/2$ is an integer (which avoids carrying around cumbersome factors of one half).
We may now expand $\left[\elliptic 00Z{\tau}\right]^{q}$ in terms of
$\sum_{p}\alpha_{p}\elliptic{\frac{p}{q}}0{qZ}{q\tau}$, where we seek $\alpha_{p}$.

We may extract $\alpha_{p}$ by expanding the two wave functions mode by mode.
After some algebra one finds
\ba
\alpha_{K} & =&
e^{-\i\pi\frac{1}{q}\tau K^{2}}\sum_{\underset{k_1,\ldots,k_q}{_{\sum_{j}k_{j}=K}}}e^{\i\pi\tau\sum_jk_j^2}\nonumber\\
 & =&\sum_{\underset{\tilde{k}_{1},\ldots,\tilde{k}_{q}}{_{\sum_{j}k_{j}=K}}}e^{\i\pi\tau\sum_{j}\tilde{k}_{j}^{2}}.
\ea
The sums are over the $k_1,\ldots,k_q\in\mathbb{Z}$, constrained such that $\sum_jk_j=K$.
We further use the abbreviation $\tilde{k}_{j}=k_{j}-\frac{K}{q}$,
to show explicitly that $\alpha_{K}=\alpha_{K+q}$, and that $\alpha_{K}=\alpha_{-K}$.
From Ref.~\cite{Fremling16b} we may identify $\alpha_{K}=Z_{K}^{\left(q\right)}$, where 
\be
Z_{k}^{\left(N\right)}=\sum_{t=1}^{N-1}Z_{t}^{\left(N-1\right)}\elliptic{\frac{t}{N-1}-\frac{k}{N}}00{N\left(N-1\right)\tau},
\ee
is defined recursively  with $Z_{t}^{\left(1\right)}=1$.
The properly normalized $\alpha_{s}$, such that $\sum_{s}\left|\alpha_{s}\right|^{2}=1$, is thus 
$\alpha_{s}=\frac{Z_{s}^{\left(q\right)}}{\sqrt{W^{\left(q\right)}}}$
where $W^{\left(q\right)} =\sum_{s=1}^{q}\left|Z_{s}^{\left(q\right)}\right|^{2}$.

The absolute value of $\alpha_{s}$ in shown in Fig.~\ref{Fig6},
for $\tau=\i\tau_2$ and in the entire $\tau$-plane, for $q=2,3$.
From this we see that in the thin-torus limit, the CF state is almost in a momentum eigenstate,
since $|\alpha_{s}^{\left(q\right)}|^2=\delta_{s,0}$.
However, in the opposite thin-torus limit ($\tau_2\to0$), the state is in an equal superposition,
with $|\alpha_{s}^{\left(q\right)}|^2=\frac{1}{q}$.

\section{The Berry curvature from Momentum Mixing}\label{App:Momentum_Mixing}

In this appendix, we compute the correction to Hall viscosity due to mixing of different momentum eigenstates, and show that it vanishes in the thermodynamic limit.
We write the wave function at $\nu={n\over 2pn+1}$ as $\Psi=\sum_{s}\alpha_{s}\Psi^{(s)}$
(in this section we omit the filling factor as a subscript, and use $s$ for the momentum index),
where $\sum_{s}\left|\alpha_{s}\right|^{2}=1$, and $\Psi^{(s)}$ are the momentum projected components.
We further assume that Eq.~\ref{eq:derivative_unproj} holds for the momentum projected states, namely that
\be
  \partial_{\bar{\tau}}\Psi^{(s)} =\i{P \over 2\tau_2}\label{eq:Bar_psi}\Psi^{(s)},\quad\quad
  \partial_{\tau}\Psi^{(s) *} =-\i{P \over 2\tau_2}\Psi^{(s) *},
  \ee
  where $P={(n+2p)N\over 4}$.
  Computing the Berry connection for the Fourier-expanded states gives
  \begin{align*}
    A_{\bar{\tau}} & =\i\big\langle\Psi\big|\partial_{\bar{\tau}}\Psi\big\rangle
    =\i\big\langle\Psi\big|\partial_{\bar{\tau}}\sum_{s}\alpha_{s}\Psi^{(s)}\big\rangle\\
    & =\i\big\langle\Psi\big|\sum_{s}\left(\partial_{\bar{\tau}}\alpha_{s}\right)\Psi^{(s)}\big\rangle
    +\i\overbrace{\big\langle\Psi\big|\sum_{s}\alpha_{s}\left(\partial_{\bar{\tau}}\Psi^{(s)}\right)\big\rangle}^{\frac{\i P}{2\tau_{2}}},
  \end{align*}
  where the last term is simply $\frac{\i P}{2\tau_{2}}$ since $\partial_{\bar{\tau}}$ preserves the momentum label.
  Expanding the bra of the first term yields
\begin{align*}
A_{\bar{\tau}} & =\i\sum_{s^{\prime}}\alpha_{s^{\prime}}^{*}\sum_{s}\left(\partial_{\bar{\tau}}\alpha_{s}\right)\overbrace{\big\langle\Psi^{(s)^{\prime}}\big|\Psi^{(s)}\big\rangle}^{\delta_{s,s^{\prime}}}-\frac{P}{2\tau_{2}}\\
 & =\i\sum_{s}\alpha_{s}^{*}\left(\partial_{\bar{\tau}}\alpha_{s}\right)-\frac{P}{2\tau_{2}}.
\end{align*}
Similarly,
\be
A_{\tau}=\i\sum_{s}\alpha_{s}\left(\partial_{\tau}\alpha_{s}^{*}\right)-\frac{P}{2\tau_{2}}
\ee
directly follows.
 The Berry curvature is thus 
\begin{align*}
\mathcal{F}_{\tau,\bar{\tau}} & =\partial_{\tau}A_{\bar{\tau}}-\partial_{\bar{\tau}}A_{\tau}\\
& =\partial_{\tau}\left(\i\sum_{s}\alpha_{s}^{*}\left(\partial_{\bar{\tau}}\alpha_{s}\right)-\frac{P}{2\tau_{2}}\right)\\
&\quad-\partial_{\bar{\tau}}\left(\i\sum_{s}\alpha_{s}\left(\partial_{\tau}\alpha_{s}^{*}\right)-\frac{P}{2\tau_{2}}\right)\\
 & =\i\sum_{s}\left[\alpha_{s}^{*}\left(\partial_{\tau}\partial_{\bar{\tau}}\alpha_{s}\right)-\alpha_{s}\left(\partial_{\bar{\tau}}\partial_{\tau}\alpha_{s}^{*}\right)\right]-\frac{\i P}{2\tau_{2}^{2}}.
\end{align*}
 Using  
\begin{align*}
\partial_{\tau} & =\left(\partial_{\tau}\tau_{1}\right)\partial_{\tau_{1}}+\left(\partial_{\tau}\tau_{2}\right)\partial_{\tau_{2}}=\frac{1}{2}\partial_{\tau_{1}}+\frac{1}{2\i}\partial_{\tau_{2}},\\
\partial_{\bar{\tau}} & =\left(\partial_{\bar{\tau}}\tau_{1}\right)\partial_{\tau_{1}}+\left(\partial_{\bar{\tau}}\tau_{2}\right)\partial_{\tau_{2}}=\frac{1}{2}\partial_{\tau_{1}}-\frac{1}{2\i}\partial_{\tau_{2}},
\end{align*}
 we can write 
$\partial_{\tau}\partial_{\bar{\tau}} =\frac{1}{4}\partial_{\tau_{1}}^{2}+\frac{1}{4}\partial_{\tau_{2}}^{2}=\frac{1}{4}\nabla^{2}$,
 so that 
 \ba
\mathcal{F}_{\tau,\bar{\tau}} &=&\i\frac{1}{4}\sum_{s}\left[\alpha_{s}^{*}\left(\nabla^{2}\alpha_{s}\right)-\alpha_{s}\left(\nabla^{2}\alpha_{s}^{*}\right)\right]-\frac{\i P}{2\tau_{2}^{2}}\nonumber\\
 &=&\frac{1}{8}\sum_{s}{\rm Im}\left[\alpha_{s}^{*}\left(\nabla^{2}\alpha_{s}\right)\right]-\frac{\i P}{2\tau_{2}^{2}}.
\ea
\ba
 \label{eq:Extra_term}
\mathcal{F}_{\tau_1,{\tau_2}} &=&-2\i \mathcal{F}_{\tau,\bar{\tau}}\nonumber\\
&=&-\frac{\i}{4}\sum_{s}{\rm Im}\left[\alpha_{s}^{*}\left(\nabla^{2}\alpha_{s}\right)\right]-\frac{P}{\tau_{2}^{2}}.
\ea
We see here that the first term in Eq.~\ref{eq:Extra_term} is of order 1, while the second term is proportional to $N$.
This means that in the thermodynamic limit $N\to\infty$, the first term is sub-leading and can be dropped.
This is why, for large systems, it does not really matter whether the state is a momentum eigenstate or not.

\section{Formulation of the wave functions with Haldane's modified Weierstrass Sigma function}
\label{app:Sigma_functions}

In the current article as well as our previous work~\cite{Pu17},
we express the wave functions for the $\nu=n/(2pn\pm 1)$ states in terms of $\vartheta$ functions. For completeness,
we show how we can formulate the wave functions using Haldane's modified Weierstrass sigma functions \cite{Haldane18}, which are defined as
\begin{widetext}
\be
\tilde{\sigma}(z,\Lambda)={\rm{exp}}\left(-{1\over 2}\gamma_2(\Lambda)z^2\right)z\prod_{L_{m,n}\neq 0}\left(1-{z\over L_{m,n}}\right){\rm exp}\left({z\over L_{m,n}}+{z^2\over 2L_{m,n}^2}\right),
\ee
\end{widetext}
where
\ba
\gamma_2(\Lambda)&=&\Gamma_2(\Lambda)-{\pi\bar{L}_1\over VL_1}\\ \nonumber
&=&\sum_{L_{m,n}\neq 0}{1\over L_{m,n}^2}-{\pi\bar{L}_1\over VL_1}.
\ea
and $V$ is the area of the unit cell.
As the definition contains all lattice vectors $L_{m,n}=mL_1+nL_2$,
$\tilde{\sigma}(z,\Lambda)$ is independent of the basis or the modular parameter $\tau$.
The modified Weierstrass sigma function has the following periodic property:
\be
{\tilde{\sigma}(z+L_i)\over \tilde{\sigma}(z)}=-{\rm{exp}}\left({\pi \over V}\bar{L}_i(z+L_i/2)\right) \quad i=1,2,
\ee
.
Given the symmetry between the $L_1$ and $L_2$ directions,
this building block is especially useful for the symmetric gauge.
If we choose $L_1$ to be real,
the modified Weierstrass sigma function can be converted into the theta function as 
\be
\label{s-t2}
\tilde{\sigma}(z,\Lambda)\propto \rm{exp}\left({z^2\over 4N_\phi \lm^2}\right)\elliptic{{1\over 2}}{{1\over 2}}{z\over L_1}{\tau}.
\ee
We use $\propto$ rather than $=$ because we have omitted a constant factor.

In this section we choose the symmetric gauge $\vec{A}=\frac{1}{2}B\vec{r}\times\hat{\vec{z}}$,
which generates a magnetic field $\vec{B}=-B\hat{\vec{z}}$. 
The quasi-periodic boundary conditions are still given by Eq.~\ref{pbc}.
The magnetic translation operator $t(\xi)$ in symmetric gauge is given by
\be
\label{app:magnetic translation operator}
t(\xi)=e^{-\frac{i}{2\lm^2}\hat{\vec{z}}\cdot(\vec{\xi}\times\vec{r})}T(\xi),
\ee
where $T(\xi)$ is the normal translation operator:
\be
T(\xi)h(z,\bar{z})=h(z+\xi,\bar{z}+\bar{\xi}).
\ee

We now reformulate the PWJ projected wave functions of Ref.~\cite{Pu17} in terms of $\tilde{\sigma}$.
We write the single-particle orbital in the LLL as~\cite{Haldane18}
\be
\psi^{(n)}_0(z,\bar{z})=e^{-{|z|^2\over 4\lm^2}}f^{(n)}_0(z)\quad n=0,1,2\dots N_\phi-1,
\ee
The subscript is the LL index and the superscript is the magnetic momentum index, just as in the main text.
The holomorphic part,
\be
f^{(n)}_0(z)=e^{ik^{(n)}z}\prod_{\mu=1}^{N_\phi}\tilde{\sigma}(z-w_\mu^{(n)}),
\ee
depends in turn on the momentum vector $k^{(n)}$ and the positions $w_\mu^{(n)}$ of the $N_\phi$ zeros of the sigma function. These are given by:
\be
k^{(n)}=-{iL_1\over \lm^2}\left({\phi_2-\phi_1\bar{\tau}\over4\pi N_\phi}-{1-\bar{\tau}\over 4}-{n\bar{\tau}\over 2N_\phi}\right),
\ee
\be
w_\mu^{(n)}=L_1\left({\phi_2-\phi_1\tau\over 2\pi N_\phi}+{\tau\over 2}
+{\mu-n\tau\over N_\phi}-1-{1\over 2N_\phi}\right).
\ee
The $\psi_0^{(n)}(z,\bar{z})$'s also satisfy Eq.~\ref{T1} and Eq.~\ref{T2}.
Note that in the the above definition we have introduced in an explicit $\tau$ dependence for both $k^{(n)}$ and $w_\mu^{(n)}$.
Thus, in the Haldane formulation we trade the theta function, which depends explicitly on $\tau$,
for a sigma function that implicitly still depends on $\tau$ through the positions of the zeros.

The single-particle wave function in the second LL is given by:
\ba
\psi_1^{(n)}(z,\bar{z})&=&a^\dagger \psi_0^{(n)}(z,\bar{z})\\ \nonumber
&=&\left({\bar{z}\over 2\sqrt{2}\lm}-\sqrt{2}\lm{\partial \over \partial z}\right)e^{-{|z|^2\over 4\lm^2}}f_0^{(n)}(z)\\ \nonumber
&=&e^{-{|z|^2\over 4\lm^2}}\left({\bar{z}\over \sqrt{2}\lm}-\sqrt{2}\lm{\partial \over \partial z}\right)f_0^{(n)}(z)\\ \nonumber
&\equiv&e^{-{|z|^2\over 4\lm^2}}a_f^\dagger f_0^{(n)}(z),
\ea
where $a_f^\dagger=\left({\bar{z}\over \sqrt{2}\lm}-\sqrt{2}\lm{\partial \over \partial z}\right)$ only acts on the holomorphic part $f_0^{(n)}$.
Wave functions in higher LLs can similarly be constructed.

The wave functions of integer-filled LLs are just the Slater determinants of occupied single particle orbitals:
\be
\Psi_n[z_i,\bar{z}_i]=e^{-\sum_i|z_i|^2\over4\lm^2}{\rm det}[f_m^{(n)}(z_i,\bar{z}_i)],
\ee
analogous to Eq.~\ref{eq:Psi_n} in the main text.
The Laughlin form for the filled LLL is:
\be
\label{app:1LL}
\Psi_{1}[z_i,\bar{z}_i]=e^{-{\sum_i|z_i|^2\over4\lm^2}}F_{1}(Z)\prod_{i<j}^N\tilde{\sigma}(z_i-z_j),
\ee
with
\beq
\label{F lau}
F_{1}(Z)&=&e^{iKZ}\tilde{\sigma}(Z-W),\\ \nonumber
K&=&-{\i L_1\over 4\pi N\lm^2}\left(\pi N+\phi_2+(\pi N-\phi_1)\bar{\tau}\right),\\ \nonumber
W&=&{L_1 \over 2\pi}\left(\pi N+\phi_2+(\pi N-\phi_1)\tau\right)
\eeq
Again, notice the $\tau$ dependence on the zeros of the wave function.

The wave functions for $\nu=n/(2pn\pm 1)$ before projection to LLL are written as:
\ba
\Psi^{\rm unproj}_{n\over 2pn+1}&=&e^{-\frac{\sum_i (|z_i|^2)}{4\lm^2}}\chi_n[{f}_i(z_j)]\nonumber\\
&&\left(F_{1}(Z)\prod_{i<j}^N\tilde{\sigma}(z_i-z_j)\right)^{2p},\label{app:CF product}
\ea
\ba
\Psi^{\rm unproj}_{n\over 2pn-1}&=&e^{-(1+{2\over 2pn-1})\frac{\sum_i (|z_i|^2)}{4\lm^2}}(\chi_n[{f}_i(z_j)])^*\nonumber\\ 
&&\left(F_{1}(Z)\prod_{i<j}^N\tilde{\sigma}(z_i-z_j)\right)^{2p},\label{app:CF product2}
\ea
where $\chi_n[{f}_i(z_j)]$ has the same form as in Eq.~\ref{upro chi}.
Just as in the main text there is an unusual factor in Eq.~\ref{app:CF product2},
originating from the relation $N_\phi=N_\phi^*+2pN=-|N_\phi^*|+2pN$ for $\nu={n\over 2pn-1}$.

The LLL projection of these wave functions requires LLL projection of the product of single-particle wave functions of the type $\psi_n \psi'_0$.
Following Ref.~\cite{Pu17},  the LLL projection is given by
\be
P_{\rm LLL} \psi_n \psi'_0=e^{-|z|^2\over 4\lm^2} \hat{f}_n f_0',
\ee
where \eg
\be
\hat{f}_1(z_i)=\sqrt{2}\lm^*\left[{\lm^2-\lm^{*2} \over \lm^{*2}}{\partial f_0\over \partial z_i}+{\lm^2\over \lm^{*2}} f_0 {\partial \over \partial z_i}\right],
\ee 
which has exactly the same form as Eq. 54 in Ref.~\cite{Pu17}.

It is now straightforward to apply the PWJ projection~\cite{Jain97,Jain97b} as shown in Ref.~\cite{Pu17}.
For the Jain $\nu={n\over 2pn+1}$ states, the LLL wave function is:
 \be
\label{app:CF proj}
\Psi_{n\over 2pn+1}[z_i,\bar{z_i}]=e^{-\frac{\sum_i (|z_i|^2)}{4l^2}}F_1^{2p}(Z){\chi}[\hat{g}_i(z_j)J_j^p],
\ee
where $J_i=\prod_{j(j\neq i)}\tilde{\sigma} \left(z_i-z_j \right)$ and ${\chi}[\hat{g}_i(z_j)J_j^p]$ is defined as in Eq.~\ref{chi-det}.
The $\hat{g}_m^{(n)}(z_i)$ is obtained from $\hat{f}_m^{(n)}(z_i)$ by making the replacement $\partial/\partial z_i\rightarrow 2\partial/\partial z_i$ for all derivatives acting on $J^p_i$.
For the LLL, $\hat{g}_0^{(n)}(z_i)=f_0^{(n)}(z_i)$.
For the first and second Landau levels, we have 
\be
 \label{app:2nd LL matrix element}
 \hat{g}_1^{(n)}(z)=-\frac{N_\phi-N_\phi^*}{N_\phi}\frac{\partial f_0^{(n)}(z)}{\partial z}+\frac{N_\phi^*}{N_\phi}f_0^{(n)}(z)2\frac{\partial}{\partial z},
 \ee
 and
\begin{widetext}
\be
\hat{g}_2^{(n)}(z)={(N_\phi-N_\phi^*)^2\over N_\phi^2}{\partial^2 f_0^{(n)}(z)\over\partial z^2}
-{2N_\phi^*(N_\phi-N_\phi^*)\over N_\phi^2}{\partial f_0^{(n)}(z)\over\partial z}2{\partial\over \partial z}
+{N_\phi^{*2}\over N_\phi^2}\big(2{\partial\over \partial z}\big)^2.
\ee
\end{widetext}

\section{Proof of Eq.~\ref{eq:momcom}}
\label{App:PLLL_vs_Mom_Proj}
In this appendix we prove Eq.~\ref{eq:momcom}. 
The momentum projection operator is given by\cite{Pu17}
\be
\mathcal{P}_M={1\over\sqrt{q}}\sum_{k=1}^{q}\left[e^{-i2\pi {M \over q}}t_{CM}\big({L_1\over N_\phi}\big)\right]^k,
\ee
where $q=2pn+1$ for the CF states.
It projects a $\Psi_{n\over 2pn+1}$ to eigenstates of $t_{CM}\big({L_1\over N_\phi}\big)$:
\be
t_{CM}\big({L_1\over N_\phi}\big)\mathcal{P}_M\Psi_{n\over 2pn+1}=e^{i2\pi{M\over 2pn+1}}\mathcal{P}_M\Psi_{n\over 2pn+1}.
\ee
$\mathcal{P}_M$ commutes with $(\partial_{\tau_2})_\tau$ because $\big[(\partial_{\tau_2})_\tau,t_{CM}\big({L_1\over N_\phi})\big]=0$. This can be seen by noting that in the current gauge we have
\ba
t_{CM}\big({L_1\over N_\phi}\big)&=&\prod_{i=1}^Ne^{{1\over N_\phi}{\partial\over\partial{\theta_{1,i}}}},\\
t_{CM}\big({L_2\over N_\phi}\big)&=&\prod_{i=1}^Ne^{{1\over N_\phi}\big({\partial\over\partial{\theta_{2,i}}}+i2\pi N_\phi\theta_{1,i}\big)}.
\ea
These are explicitly independent of $\tau_1$ and $\tau_2$, and hence commute with $\partial_{\tau_2}$. Therefore, Eq.~\ref{eq:momcom} does hold.

Here we emphasize that $\mathcal{P}_M$ is not modular invariant, 
as noted in the main text, even though there is no $\tau$ in its definition.
The reason is that the reduced coordinates used in the definition of $\mathcal{P}_M$ do change under modular transformations.
For instance, under $\tau\to -{1\over \tau}$, to keep the physical coordinates $(x,y)$ fixed,
the reduced coordinates transform as $(\theta_1,\theta_2)\to (-\theta_2,\theta_1)$.

\section{Direct evaluation of the correction term}
\label{app:last}

\begin{figure}[t]
  \includegraphics[width=\columnwidth]{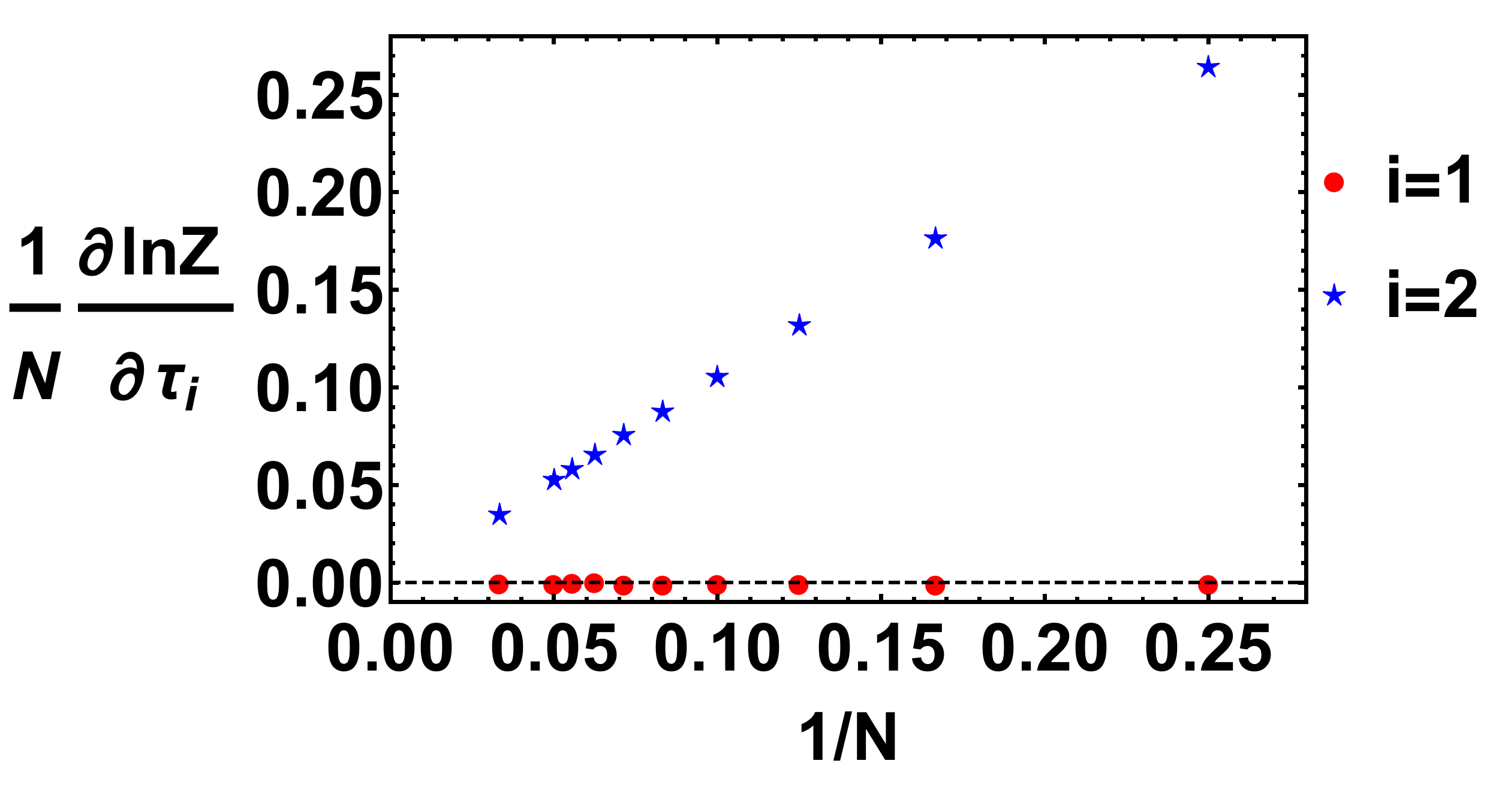} 
  \caption{
    ${1\over N}\left({\partial \\ln  Z\over \partial \tau_1}\right)_{\tau_2}$ (red dots) and
    ${1\over N}\left({\partial \\ln  Z\over \partial \tau_2}\right)_{\tau_1}$ (blue stars)
    are depicted for the unprojected $\nu=2/5$ wave function as a function of $1/N$.  Both clearly vanish in the thermodynamic limit. The calculations are performed for  $\tau=\i$.
  }
  \label{Fig7}
\end{figure}

Our analytical derivation of the Hall viscosity of the unprojected Jain wave functions is based on the assumption that the overall normalization factor satisfies Eq.~\ref{Z_infty}. For a numerical test, we first write
\be
\label{norm assum2}
{1\over N}\left({\partial \ln Z\over \partial \tau_2}\right)_\tau
=-\i{1\over N}\left({\partial \ln Z\over \partial \tau_1}\right)_{\tau_2}+{1\over N}\left({\partial \ln Z\over \partial \tau_2}\right)_{\tau_1}.
\ee
We numerically evaluate the two terms on the right-hand side for $\nu=2/5$. The results, displayed in Fig.~\ref{Fig7},
show that both ${1\over N}\left({\partial \ln Z\over \partial \tau_1}\right)_{\tau_2}$ and
${1\over N}\left({\partial \ln Z\over \partial \tau_2}\right)_{\tau_1}$ vanish in the thermodynamic limit.

\begin{figure}[t]
  \includegraphics[width=\columnwidth]{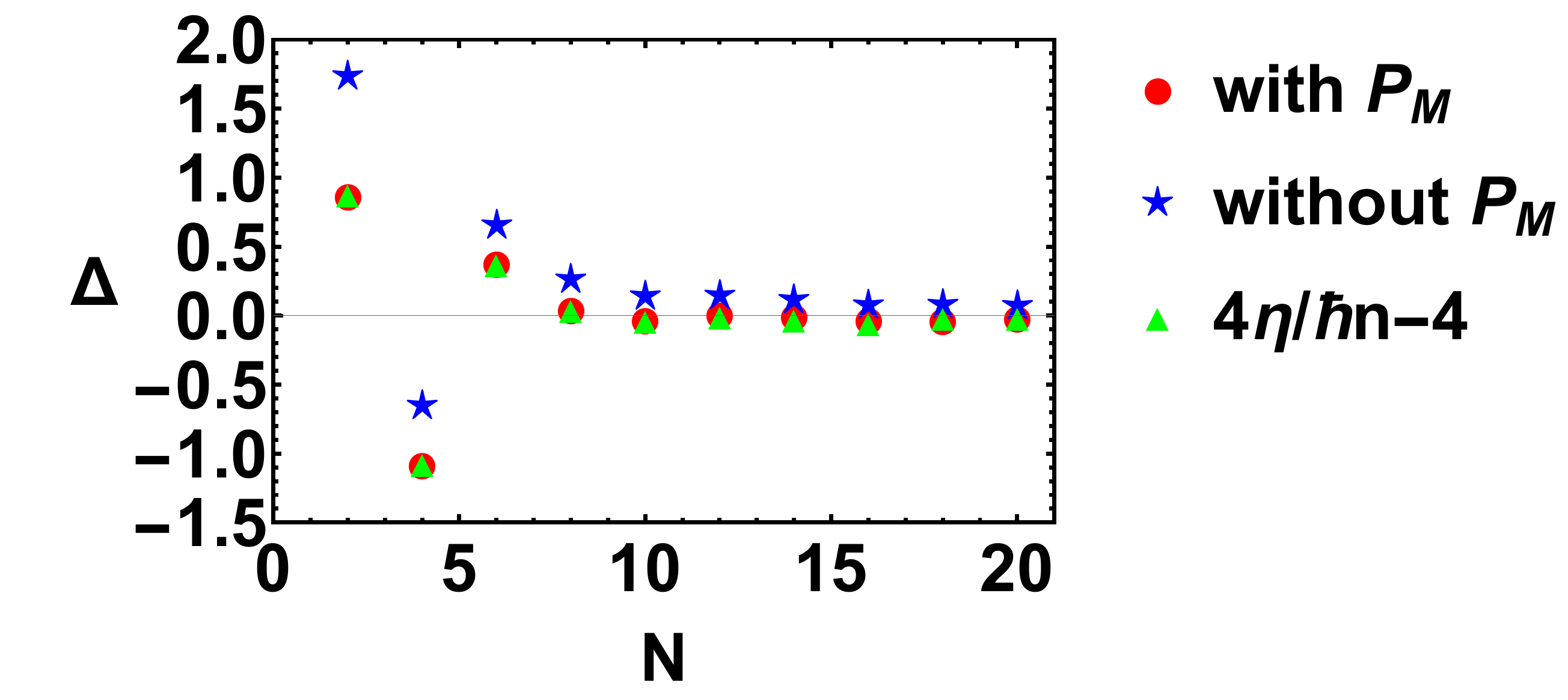} 
  \caption{
    The numerical results of the correction $\Delta$ at $\tau=\i$ for $\nu=2/5$,
    for momentum-projected and unprojected wave functions.
    As a cross-check the actual difference $\Delta={4\eta V\over \hbar N}-4$, with $\eta$ taken from Fig.~\ref{Fig1}, is also included.
  }
  \label{Fig8}
\end{figure}

We can explicitly evaluate the contribution of this term to Hall viscosity.
Without assuming the condition in Eq.~\ref{Z_infty}, Eq.~\ref{eq:derivative_unproj} becomes
\be
\label{eq:derivative_unproj2}
(\partial_{\tau_2})_\tau \Psi_{n\over 2pn+1}^{\rm unproj}=\left[(\partial_{\tau_2})_\tau \ln Z+{(n+2p)N\over 4\tau_2}\right]\Psi_{n\over 2pn+1}^{\rm unproj}
\ee
Let us denote 
\be
\gamma\equiv (\partial_{\tau_2})_\tau \ln Z=\left({\partial \ln Z\over\partial \tau_2}\right)_{\tau_1}-\i \left({\partial \ln Z\over\partial \tau_1}\right)_{\tau_2}.
\ee
Following the calculation in Sec.~\ref{sec:Analytical_proof}, we get
\be
A_\tau=-{1\over 2}\gamma^*-{1\over 2}{N(n+2p)\over 4\tau_2},
\ee
\be
A_{\bar{\tau}}=-{1\over 2}\gamma-{1\over 2}{N(n+2p)\over 4\tau_2},
\ee
\be
A_1=A_\tau+A_{\bar{\tau}}=-{\rm Re} \gamma-{N(n+2p)\over 4\tau_2},
\ee
and
\be
A_2=\i(A_\tau-A_{\bar{\tau}})=-{\rm Im} \gamma.
\ee
Finally, the Berry curvature is
\ba
\mathcal{F}_{\tau_1,\tau_2}&=&N\left[-{n+2p\over 4\tau_2^2}+{1\over N}\left(-\partial_{\tau_1}{\rm Im}\gamma+\partial_{\tau_2}{\rm Re}\gamma\right)\right]\nonumber \\
&=&N\left[-{n+2p\over 4\tau_2^2}+{1\over N}\left((\partial_{\tau_1})^2\ln Z+(\partial_{\tau_2})^2\ln Z\right)\right]\nonumber \\
&=&-N{n+2p+\Delta \over 4\tau_2^2},
\ea
where
\ba
\label{Delta}
\Delta&\equiv& -{4\tau_2^2\over N}\left((\partial_{\tau_1})^2\ln Z+(\partial_{\tau_2})^2\ln Z\right)\nonumber\\
      &=&{4\eta V \over \hbar N}-(n+2p).
\ea

In Fig.~\ref{Fig8}, we show the evaluation of $\Delta$ for wave functions both before and after momentum projection.
(They have different overall normalization factors $Z$.)
We also show, as a sanity check, the actual difference ${4\eta\over \hbar n}-4$ using the results from Fig.~\ref{Fig1}.

As shown in Fig.~\ref{Fig8}, the correction term converges to zero quickly with increasing system sizes,
especially for the normalization factor after momentum projection.
The correction $\Delta$ also agrees with the difference between ${4\eta V\over \hbar N}$ which is obtained in Fig.~\ref{Fig1} and the quantized shift $\sh=4$ for $\nu={2\over 5}$.

Although we do not know analytically how the convergence to the thermodynamic limit occurs with increasing system size, we can give the following argument. When the system is large enough that the individual composite fermion does not know the global shape of the whole system, the dependence of $\ln Z$ on $\tau_1$ and $\tau_2$ is unlikely to be extensive. Therefore, according to Eq.~\ref{Delta} the scaling behavior of $\Delta$ with $N$ is expected to be $\ln \Delta=-\alpha\ln N$ with $\alpha>1$, i.e., the correction should converge to zero no slower than $1/N$. Indeed, we estimate $\alpha \approx 5$ for the data shown in Fig.~\ref{Fig8} for system sizes between 6 and 12 particles (larger systems are not included because the error bars become comparable to the deviation $\Delta$ itself). This qualitative argument is also supported by Fig.10 of Ref.~\cite{Fremling16b}.

\end{appendix}

\bibliographystyle{apsrev}

\end{document}